This figure "fig1-17.png" is available in "png" format from:

http://arxiv.org/ps/hep-lat/9402011v1

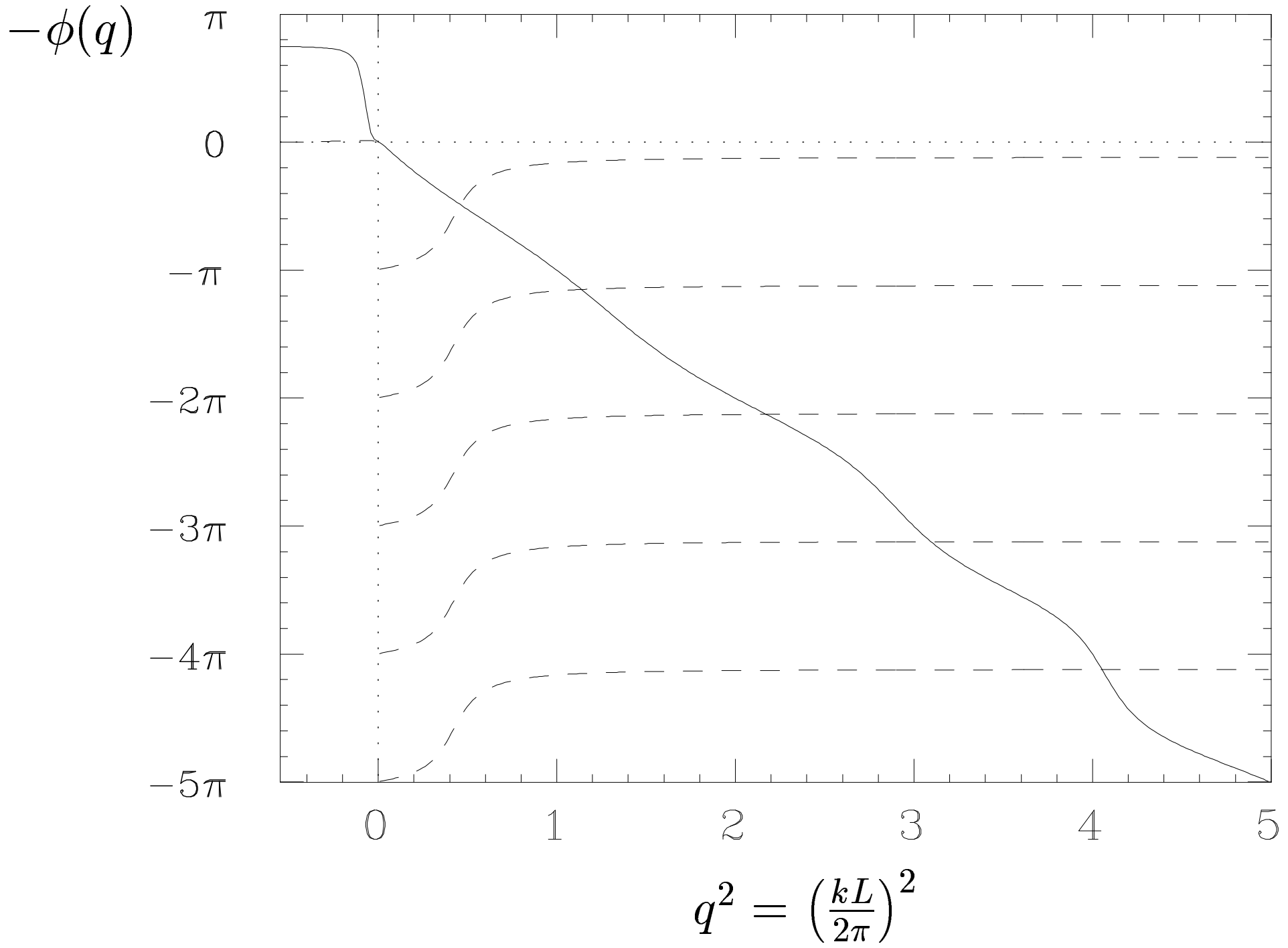

**Figure 1**: The functions $-\phi(q)$ (solid line) and $\delta_0(k) - \nu\pi$ for $\nu = 0, 1, 2, \ldots$ (dashed lines) for a fixed value of $L$. $\delta_0(k)$ is calculated perturbatively according to eq. (3.6).

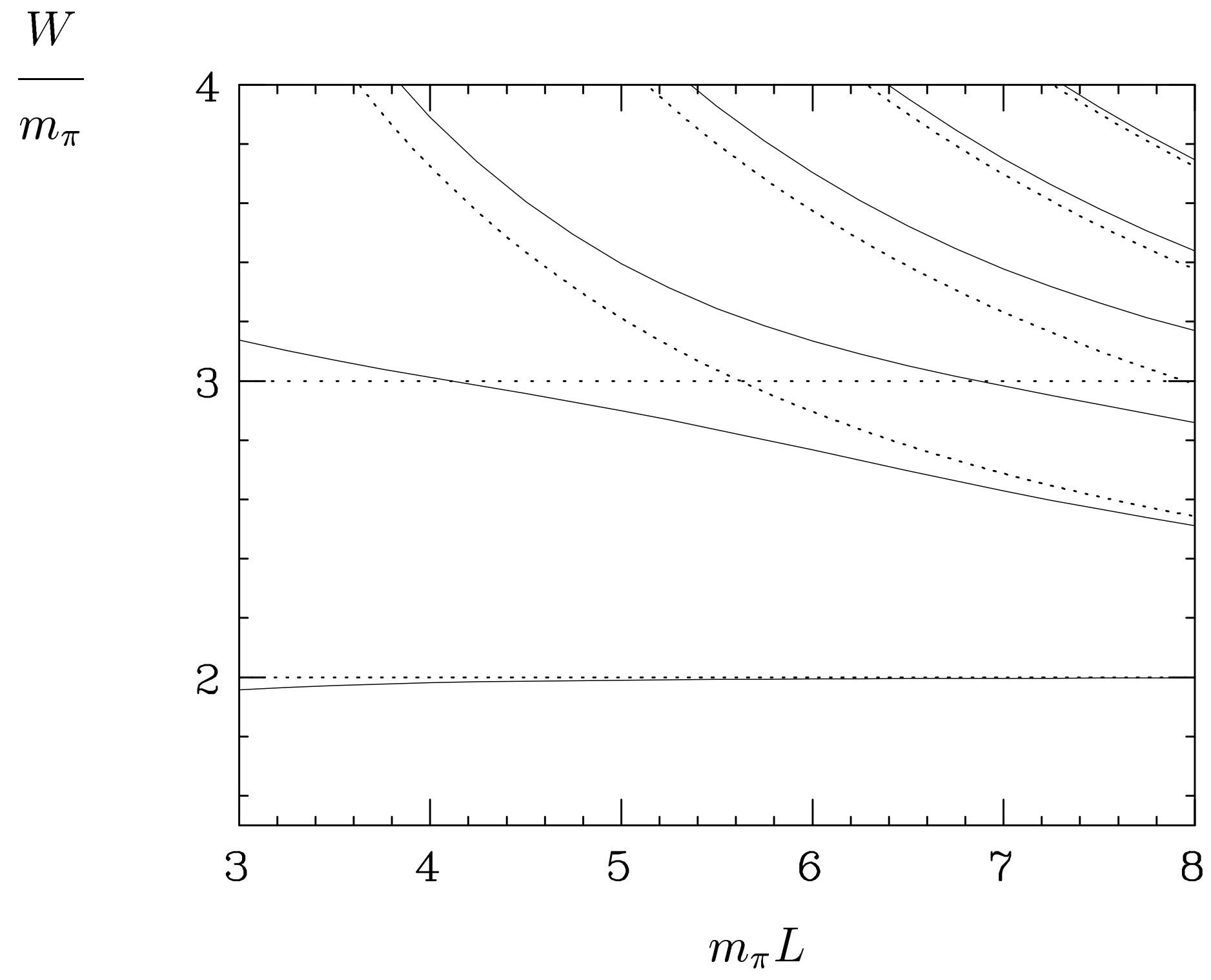

**Figure 2**: Example of a perturbatively predicted two-particle energy spectrum for isospin 0 (solid lines) with a value $g_R = 24$ for the renormalized coupling and a mass ratio $m_\sigma/m_\pi = 3$ (with $m_\pi = 0.23$). The dotted lines show the free energy spectrum (3.9). The $\sigma$-resonance energy $m_\sigma$ is marked by the horizontal line at $W = 3m_\pi$.

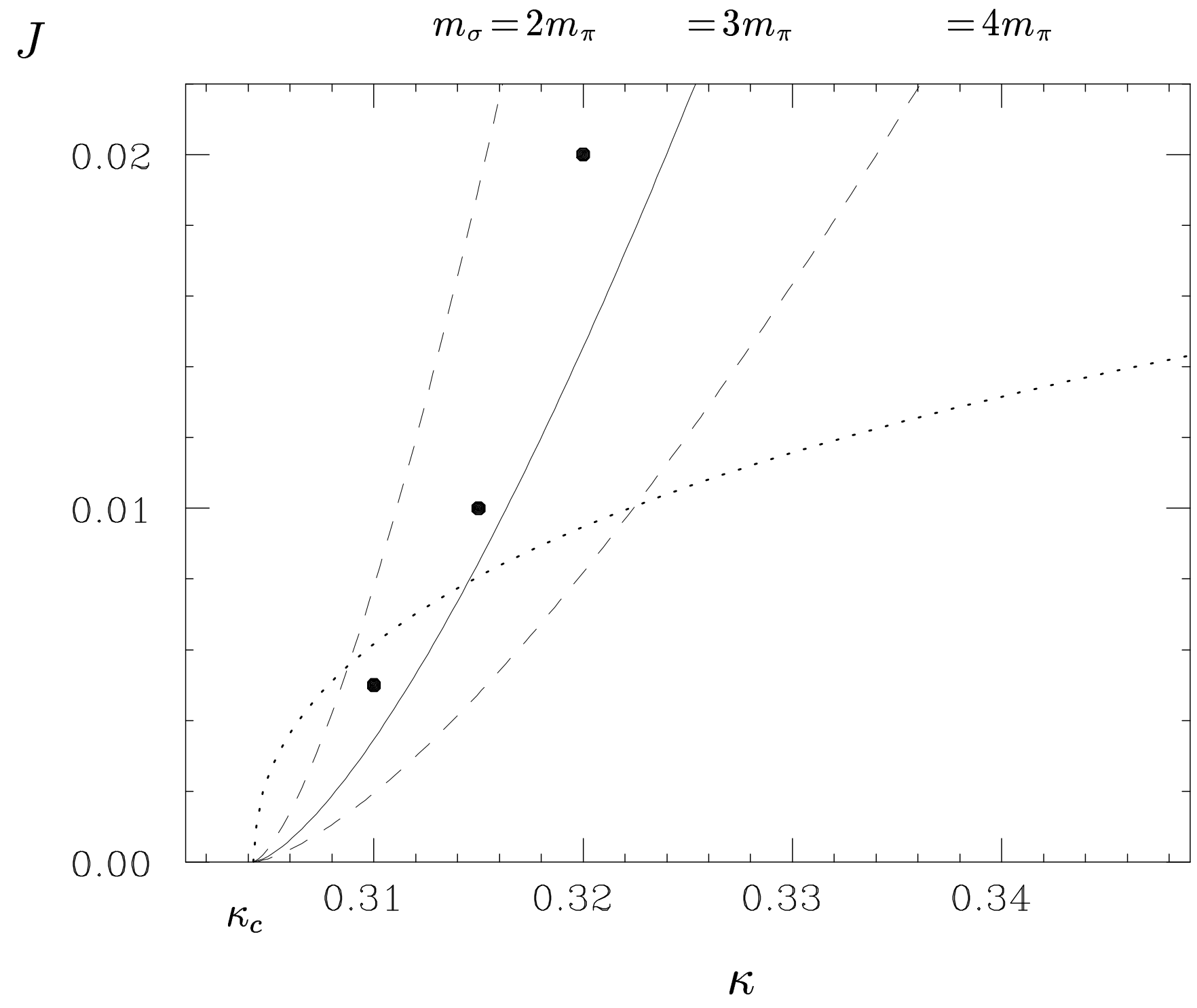

**Figure 3**: Lines of constant mass ratios $m_\sigma/m_\pi$ and $m_\pi L$ in the $\kappa$-$J$ plane derived with the help of the scaling laws of ref. [22]. The solid line represents the ratio $m_\sigma/m_\pi = 3$. Our choice of parameters ($\kappa$, $J$) near that line is symbolized by the thick dots. To the left and to the right, the region where elastic two-particle scattering is expected is bounded by dashed lines corresponding to $m_\sigma/m_\pi = 2$ and $m_\sigma/m_\pi = 4$, respectively. The dotted line corresponds to $m_\pi L = 3$ for $L = 16$.

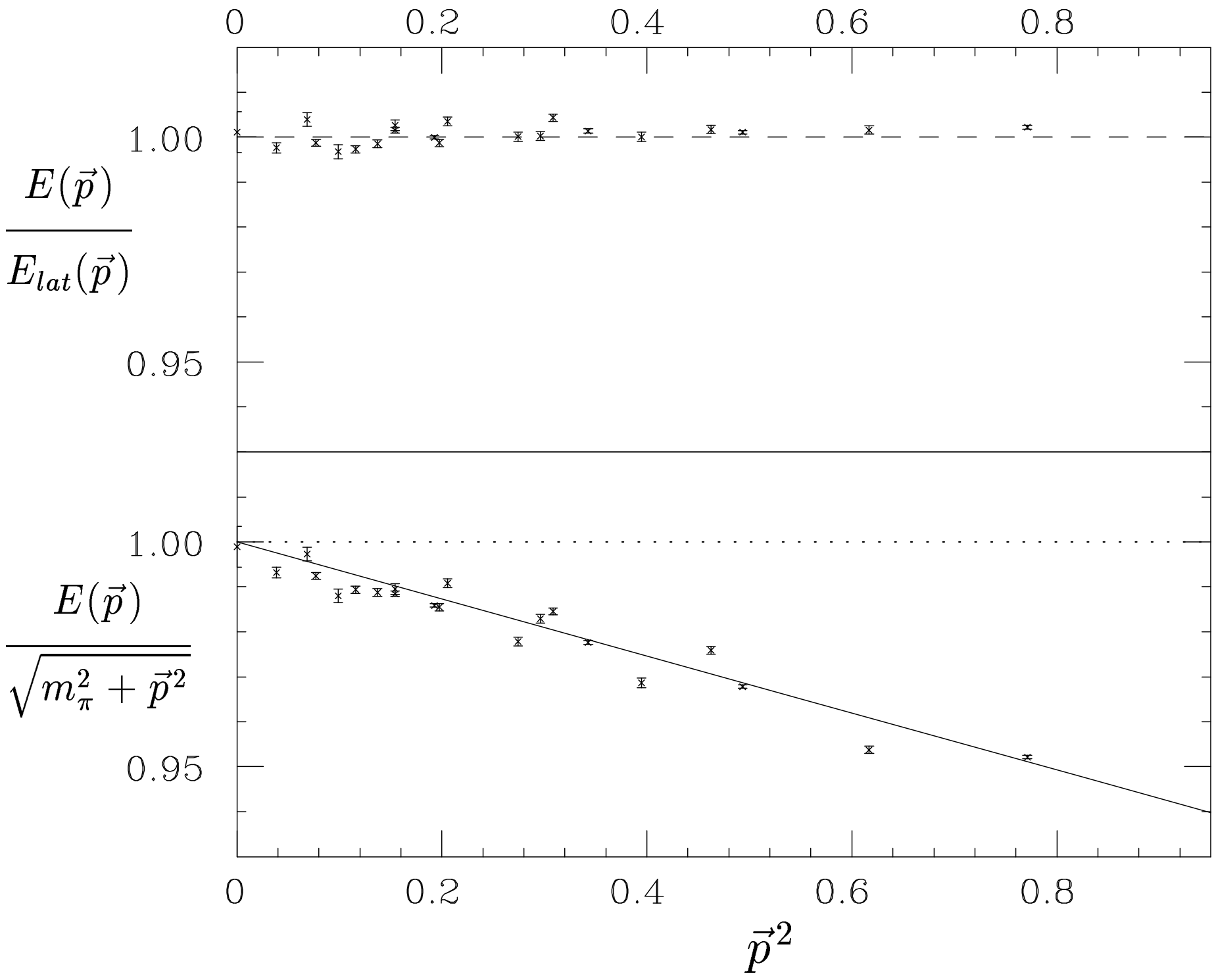

**Figure 4**: Single pion energy values $E(\vec{p})$ divided by the lattice and the continuum dispersion relation, respectively, as function of $\vec{p}^2$ at the simulation point ($\kappa = 0.315$, $J = 0.01$).

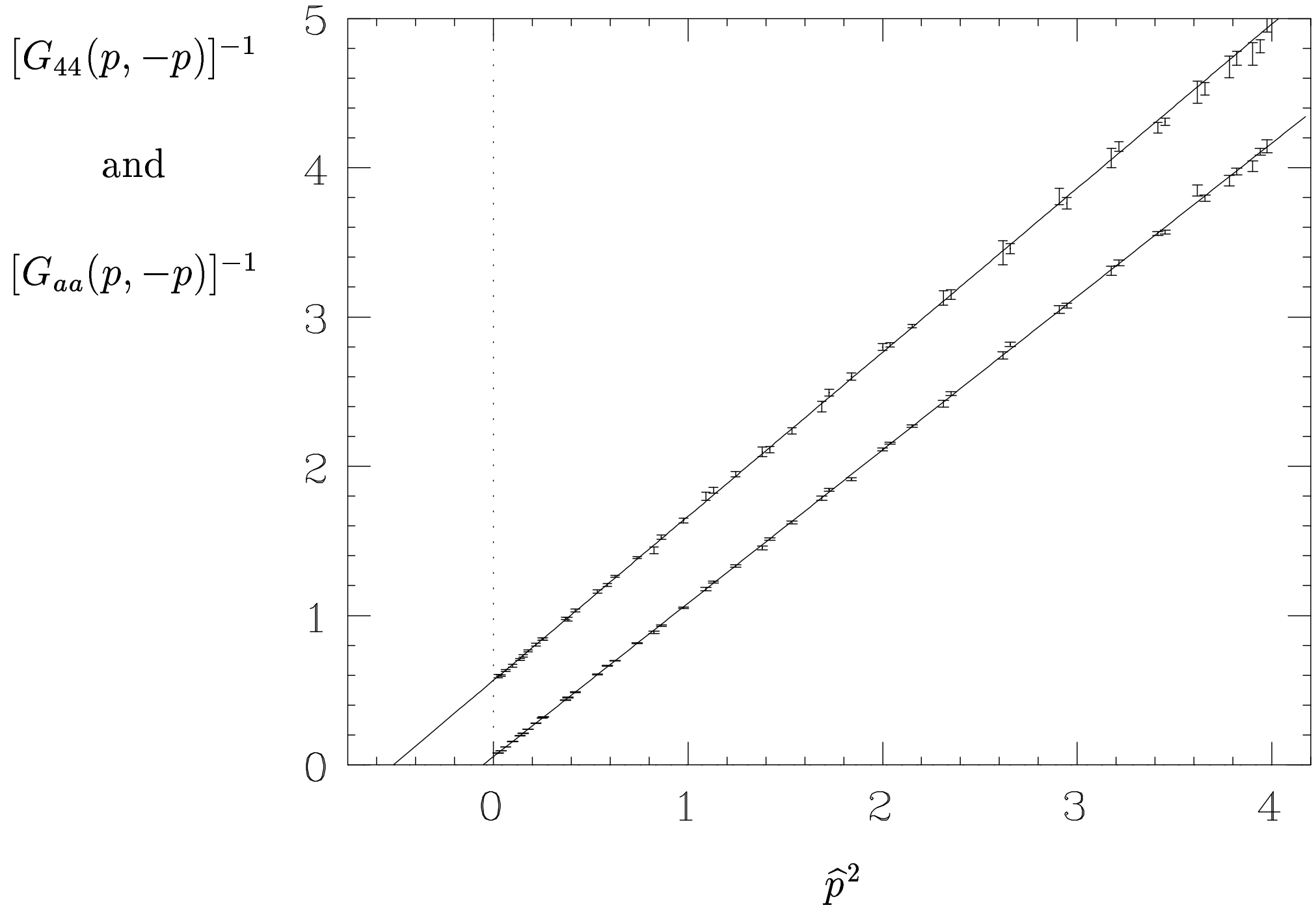

**Figure 5**: The upper (lower) curve shows a two-parameter fit to the inverse $\sigma$ ($\pi$) propagator in momentum space. The data are for the $32^3 \cdot 40$-lattice at ($\kappa = 0.315$, $J = 0.01$). The intercept with the abscissa yields $-\tilde{m}_\sigma^2$ ($-m_\pi^2$).

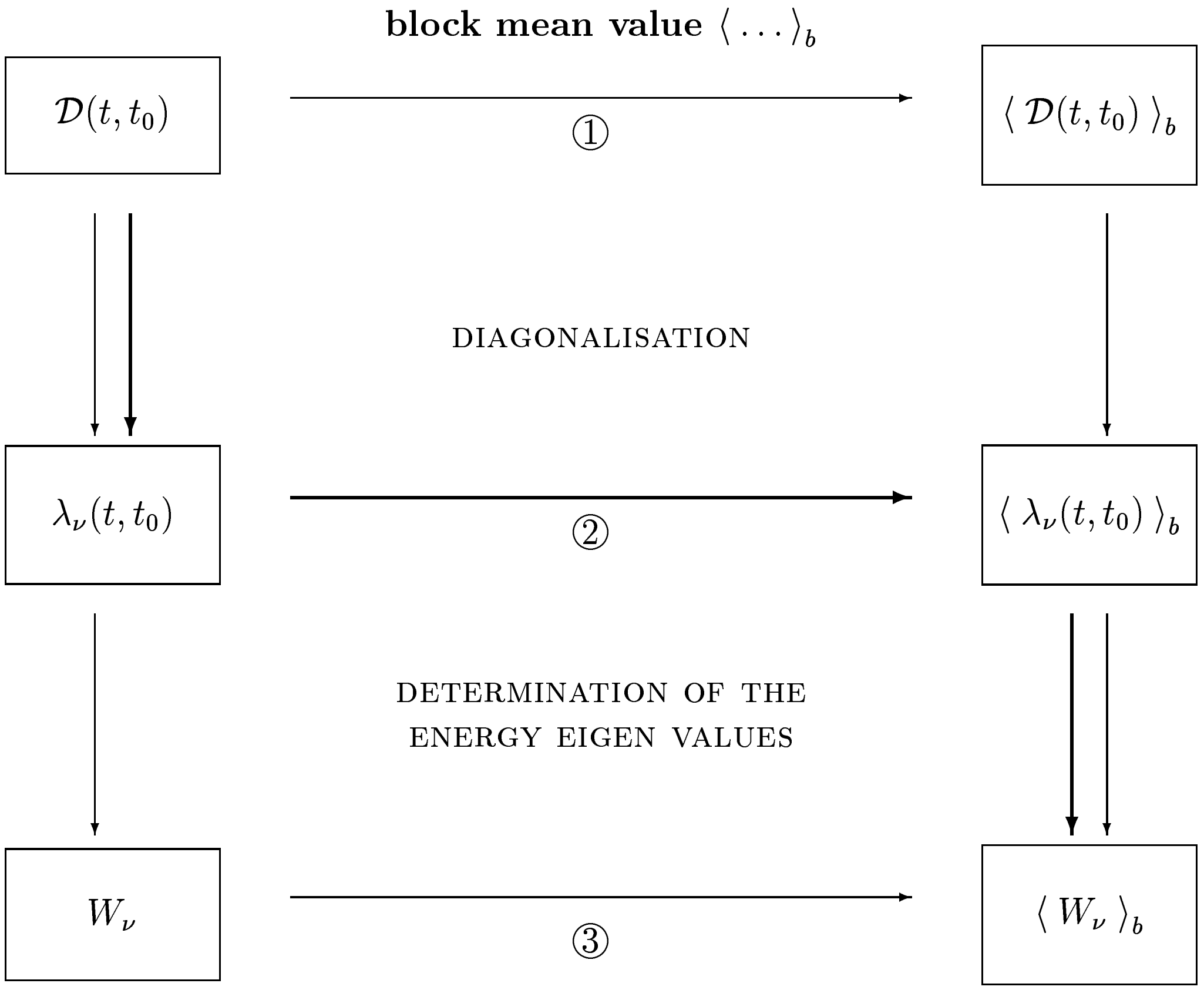

**Figure 6**: Different strategies for determining the block mean values of the two-particle energy spectrum.

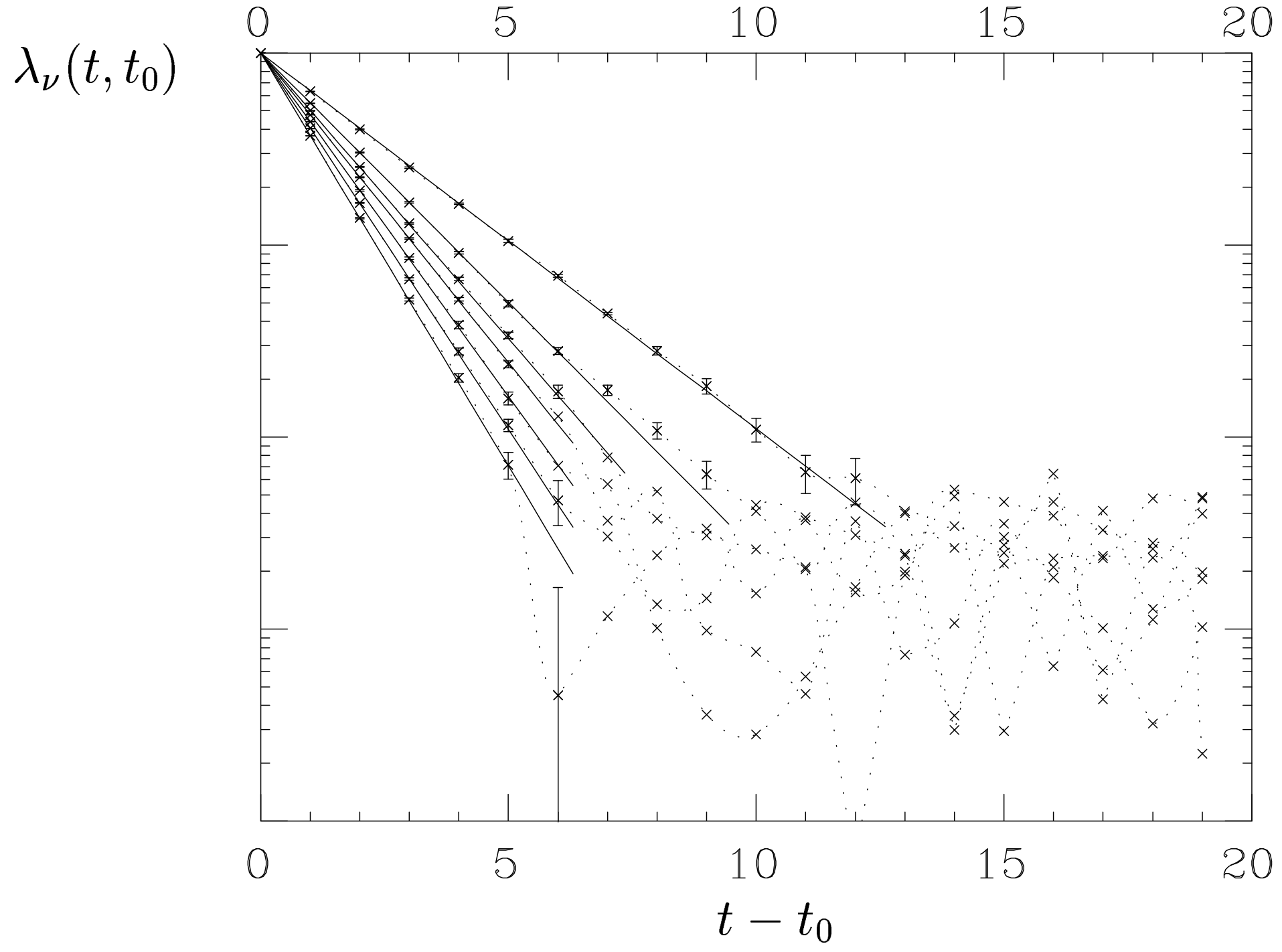

**Figure 7**: The self-adjusting fit of the (mean) eigenvalues $\lambda_\nu(t,t_0)$ of the correlation function matrix $\mathcal{D}(t,t_0)$ for isospin 0 at ($\kappa = 0.315$, $J = 0.010$, $L = 32$): The eigenvalues on the different time slices are connected by dotted lines according to the criterion of the highest possible collinearity of the eigenvectors. Crossing of lines implies a contradiction to the sorting of the eigenvalues with respect to their size, which is expected to be equivalent to the former method. The crossing point determines the upper end $t_{max}$ of the range used in the exponential fit (solid line).

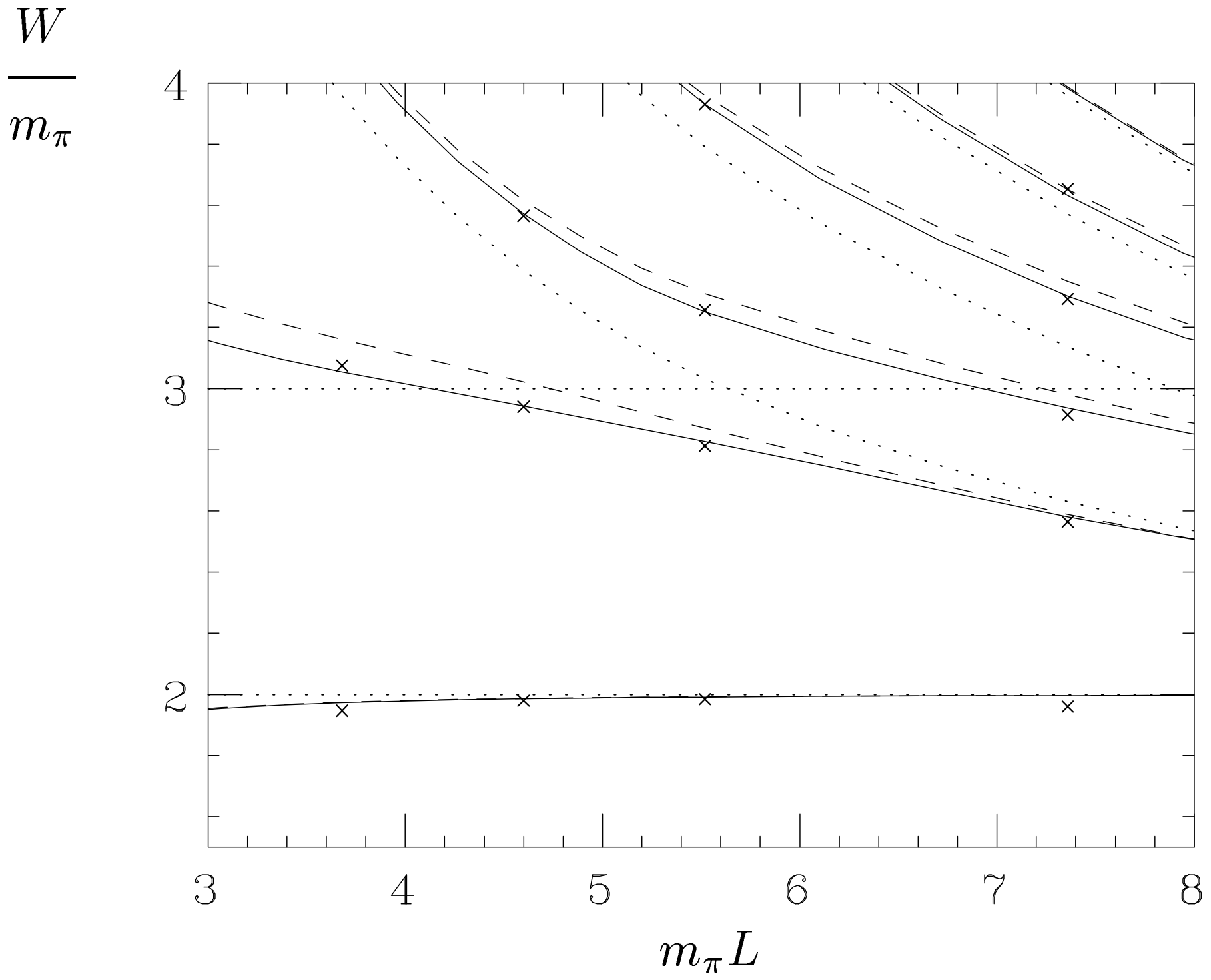

**Figure 8**: Comparison of the Monte Carlo results (crosses) with the perturbatively predicted two-particle energy spectrum in the isospin-0 channel for the simulation point ($\kappa = 0.315$, $J = 0.01$). The data are in excellent agreement with the perturbative predictions (solid lines) based on the results of table 11. Dotted lines refer to the free energy spectrum, while the dashed lines show the perturbative spectrum based on the estimates of table 4. The location of the resonance energy $m_\sigma$ is symbolized by the dotted horizontal line at $W \approx 3m_\pi$. – Errors are smaller than the symbols.

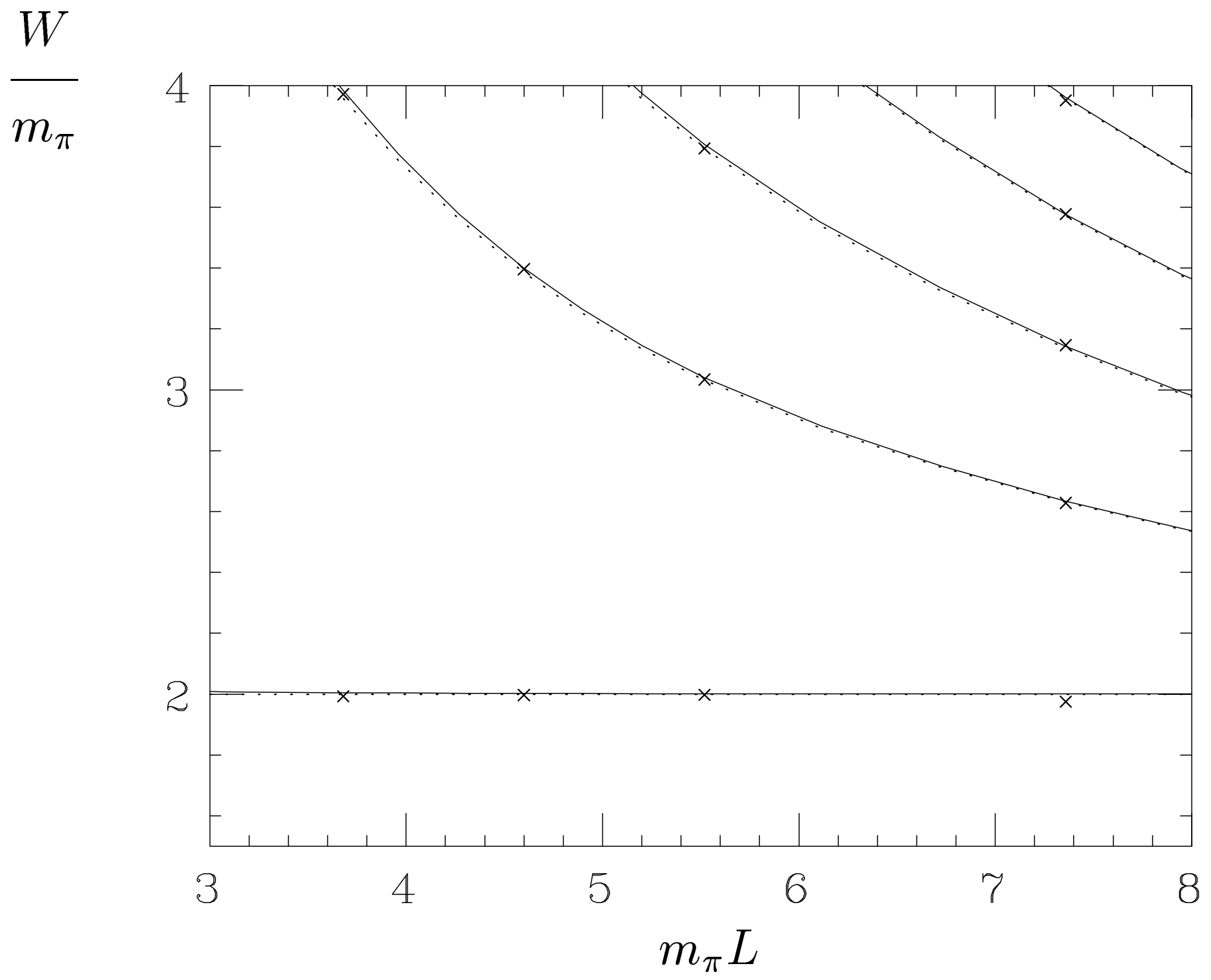

**Figure 9**: Comparison of the Monte Carlo results (crosses) with the perturbatively predicted two-particle energy spectrum in the isospin-2-channel for the simulation point ($\kappa = 0.315$, $J = 0.01$). The data are in excellent agreement with the perturbative predictions (solid lines) based on results of table 11. Calculations based on the estimates of table 4 yield indistinguishable results. Dotted lines refer to the free energy spectrum. – Errors are smaller than the symbols.

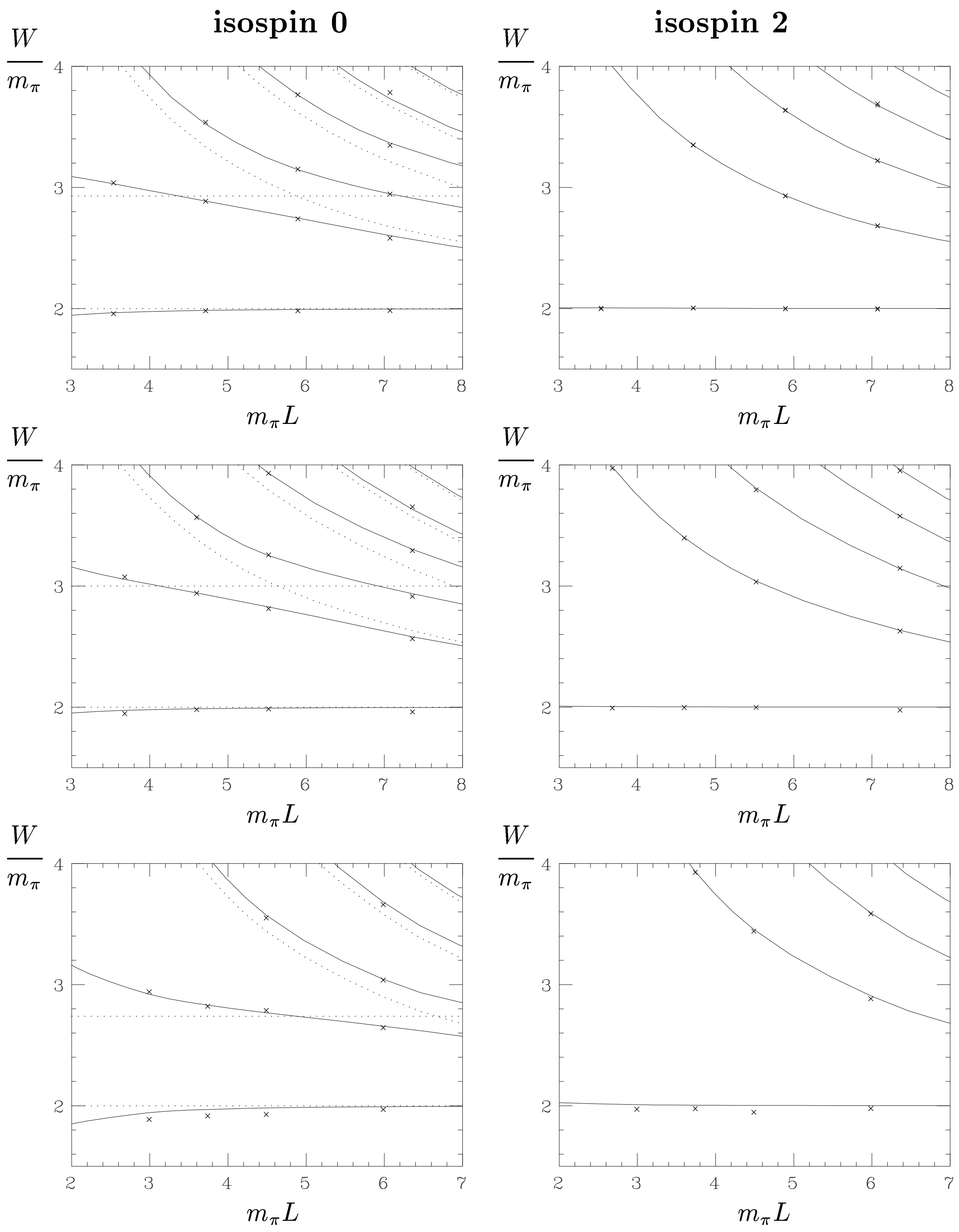

**Figure 10**: All data for the two-particle energy spectrum for isospin 0 (l.h.s.), and for isospin 2 (r.h.s.). The dotted lines in the isospin-0 case show the free spectrum. For both isospins and all simulation points ($\kappa = 0.320$, $J = 0.020$, top), ($\kappa = 0.315$, $J = 0.010$, middle) and ($\kappa = 0.310$, $J = 0.005$, bottom), the data are in excellent agreement with the perturbative calculations (solid lines) provided we use the resonance parameters of table 11 in the case of isospin 0. – Errors are smaller than the symbols.

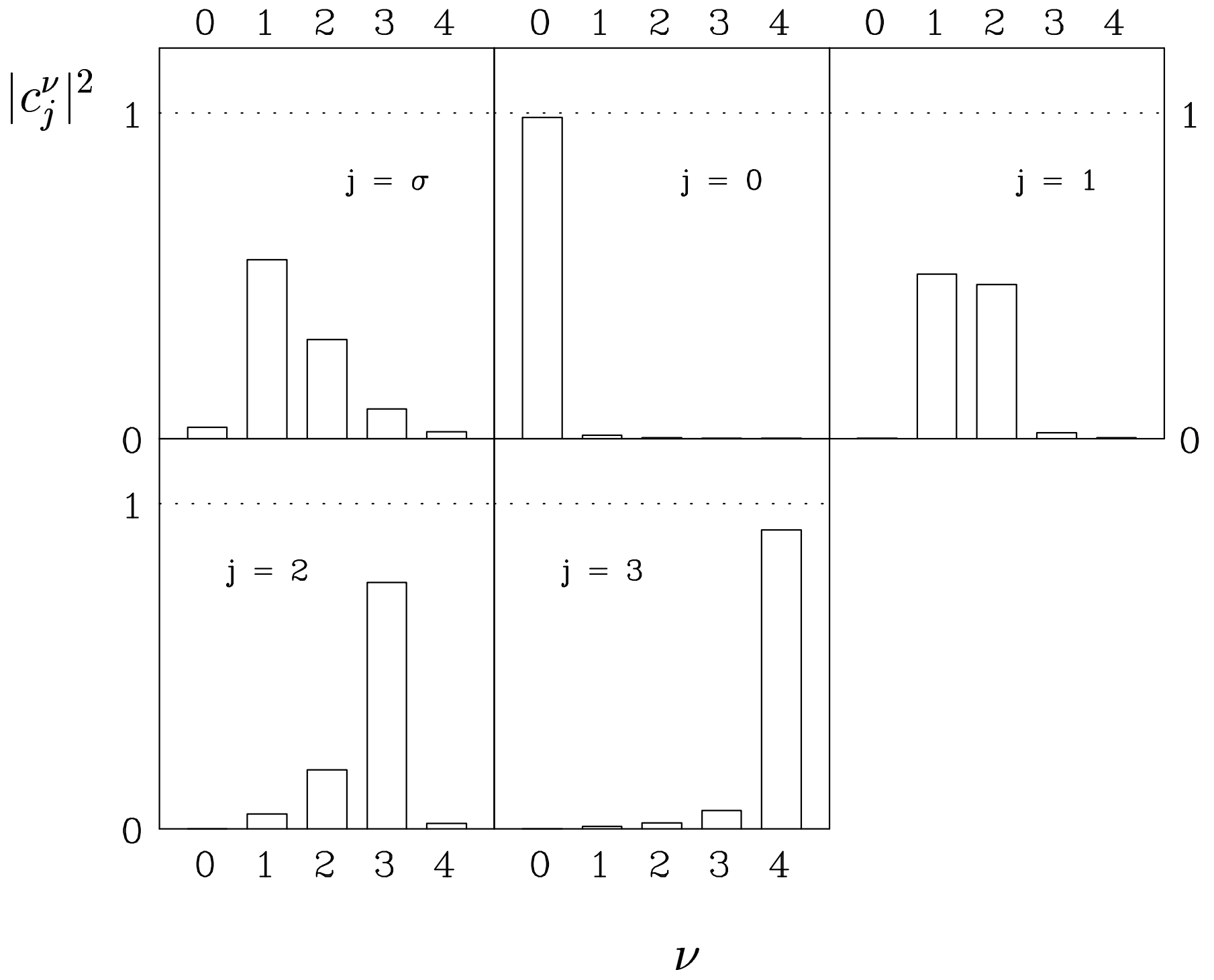

**Figure 11**: Normalized spectral amplitudes $|c_j^\nu|^2$ in the isospin-0 channel for the operators $O_j$, $j=\sigma,0,1,\ldots$ in the case of plane waves at ($\kappa=0.315$, $J=0.01$, $L=24$).

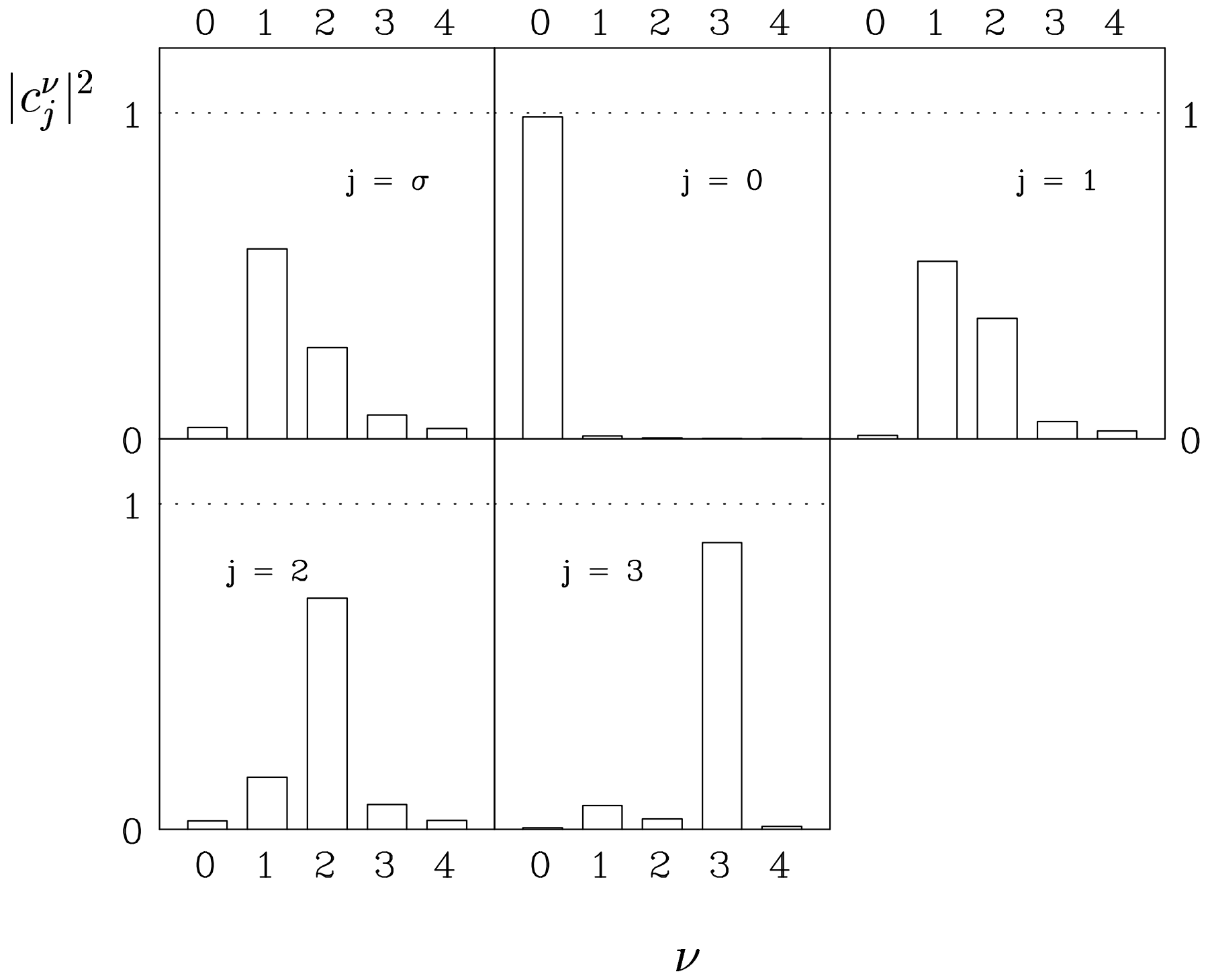

**Figure 12**: Normalized spectral amplitudes $|c_j^\nu|^2$ in the isospin-0 channel for the operators $O_j$, $j = \sigma, 0, 1, \ldots$ in the case of Lüscher's wave functions at ($\kappa\!=\!0.315$, $J\!=\!0.01$, $L\!=\!24$).

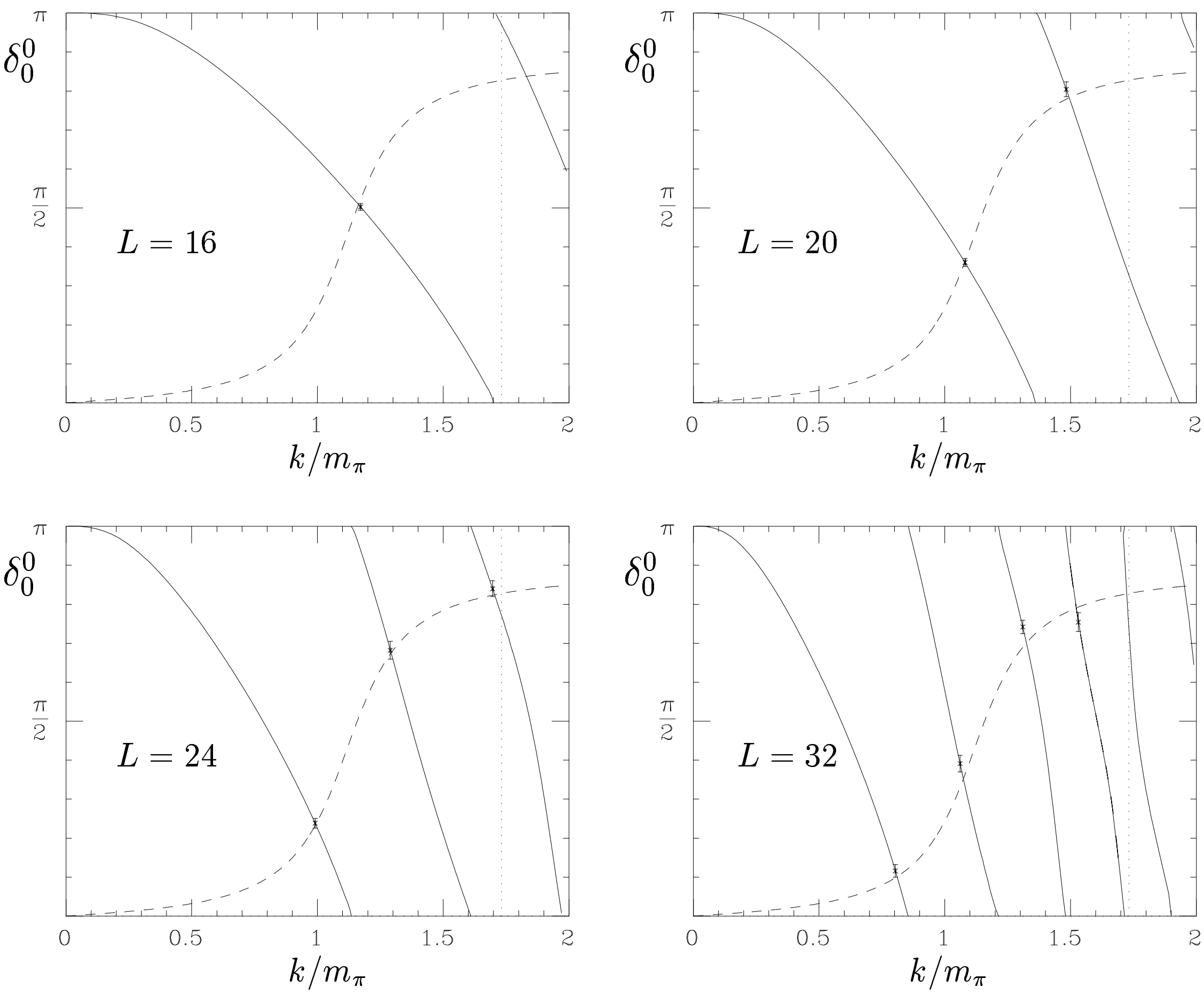

**Figure 13**: $-\phi\left(\frac{kL}{2\pi}\right) \bmod \pi$ as function of $k/m_\pi$ with $L$ fixed for $(\kappa = 0.315,\ J = 0.01)$ and $L = 16,\ 20,\ 24,\ 32$. The function $-\phi$ taken at the momentum $k$ corresponding to one of the energy values of table 8 determines the scattering phase shift at this energy. The error of the scattering phase shift is calculated from the error of $k$ using the slope of $\phi$. The dashed lines represent the behaviour of the scattering phase shift according to a fit with respect to the perturbative ansatz (3.6) (compare with fig. 1). The dotted vertical line indicates the inelastic threshold.

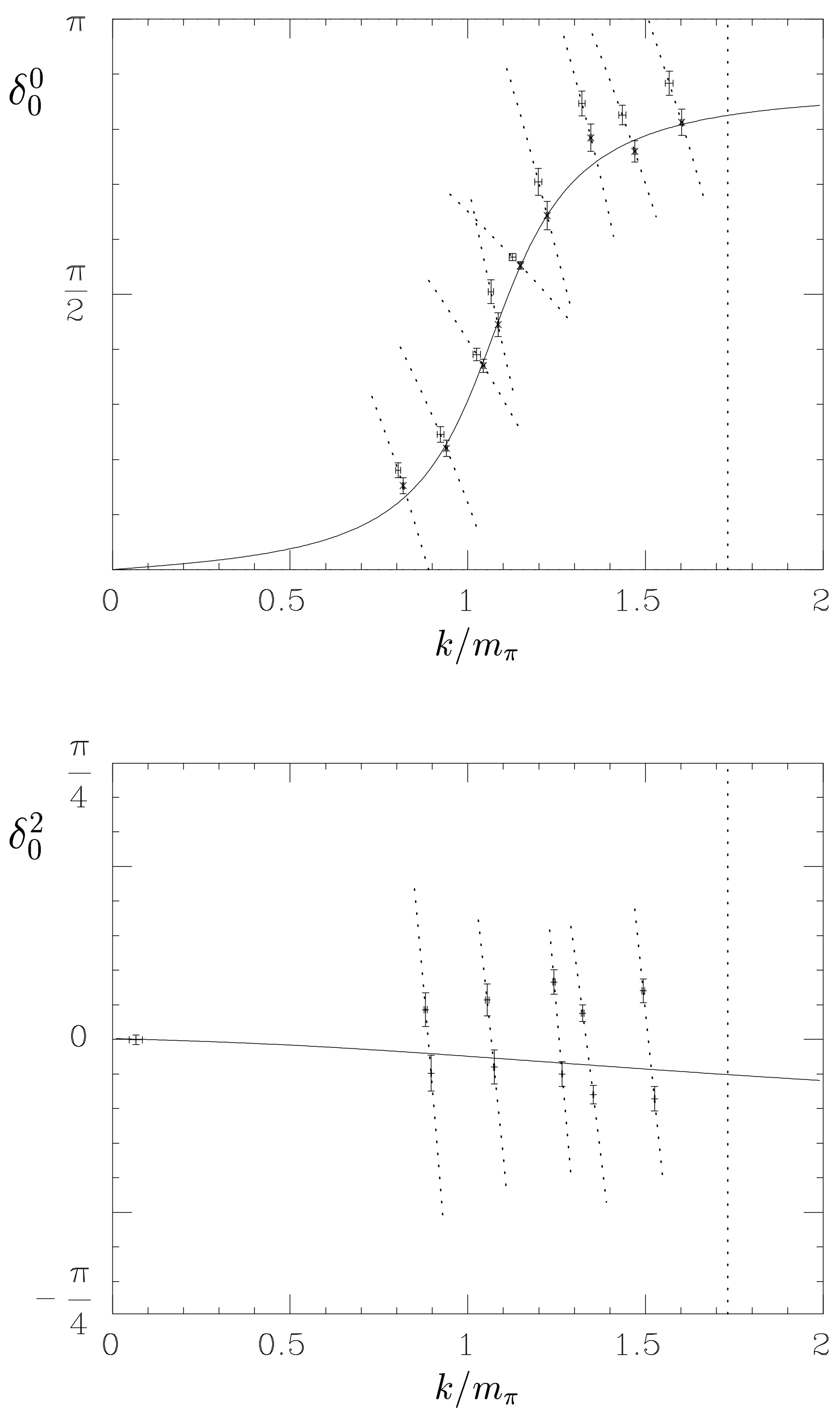

**Figure 14**: Lattice effects in the calculation of the scattering phase shifts for isospin 0 and 2: the upper (lower) points are determined by means of the continuum (lattice) dispersion relation. The scattering phases in the isospin-0 channel are well explained by a fit with the perturbative ansatz (3.6), and for isospin 2 the data are in good agreement with perturbative calculations based on these results (solid curves). Here we show the results at the data point ($\kappa = 0.320$, $J = 0.02$) with the largest lattice effects.

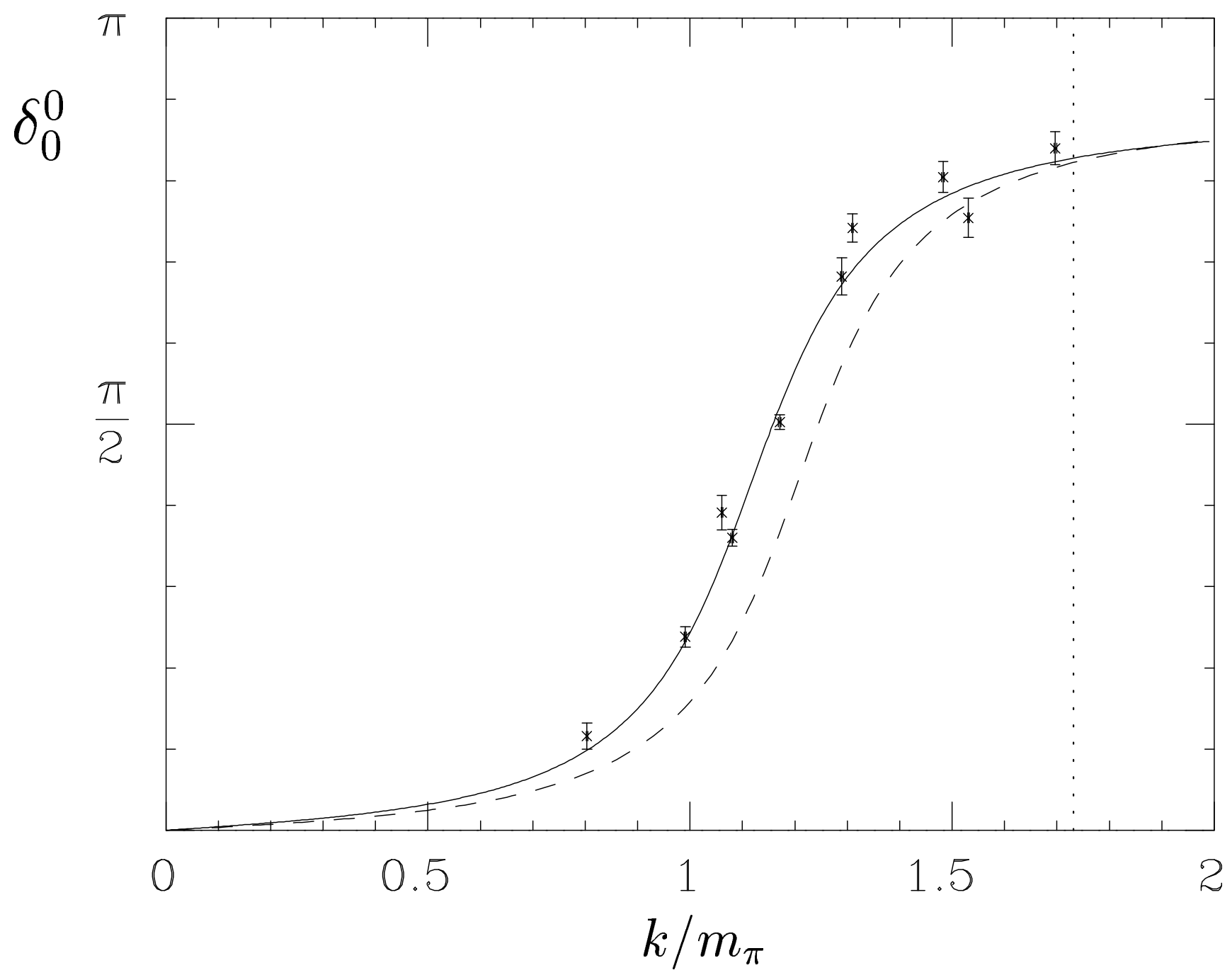

**Figure 15**: Comparison of the fit with respect to the perturbative ansatz (3.6) (solid curve) and the perturbative prediction based on the estimates $\tilde{m}_\sigma$ and $\tilde{g}_R$ of table 4 (dashed curve) in the isospin-0 channel for ($\kappa = 0.315$, $J = 0.01$).

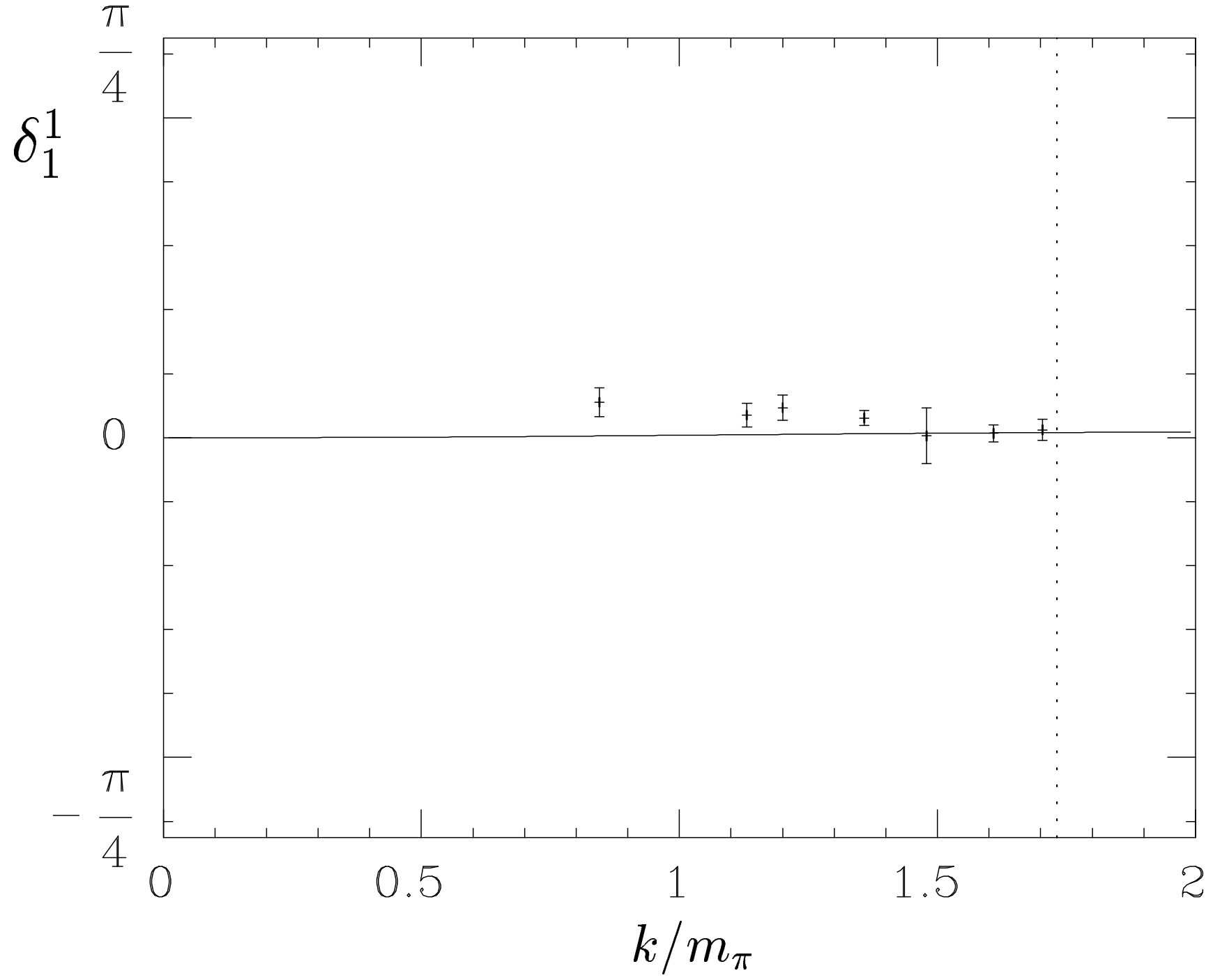

**Figure 16**: Scattering phase shifts for isospin 1 at ($\kappa = 0.315$, $J = 0.01$). The data are in good agreement with perturbative calculations based on the estimates $\tilde{m}_\sigma$ and $\tilde{g}_R$ of table 4 (solid curves).

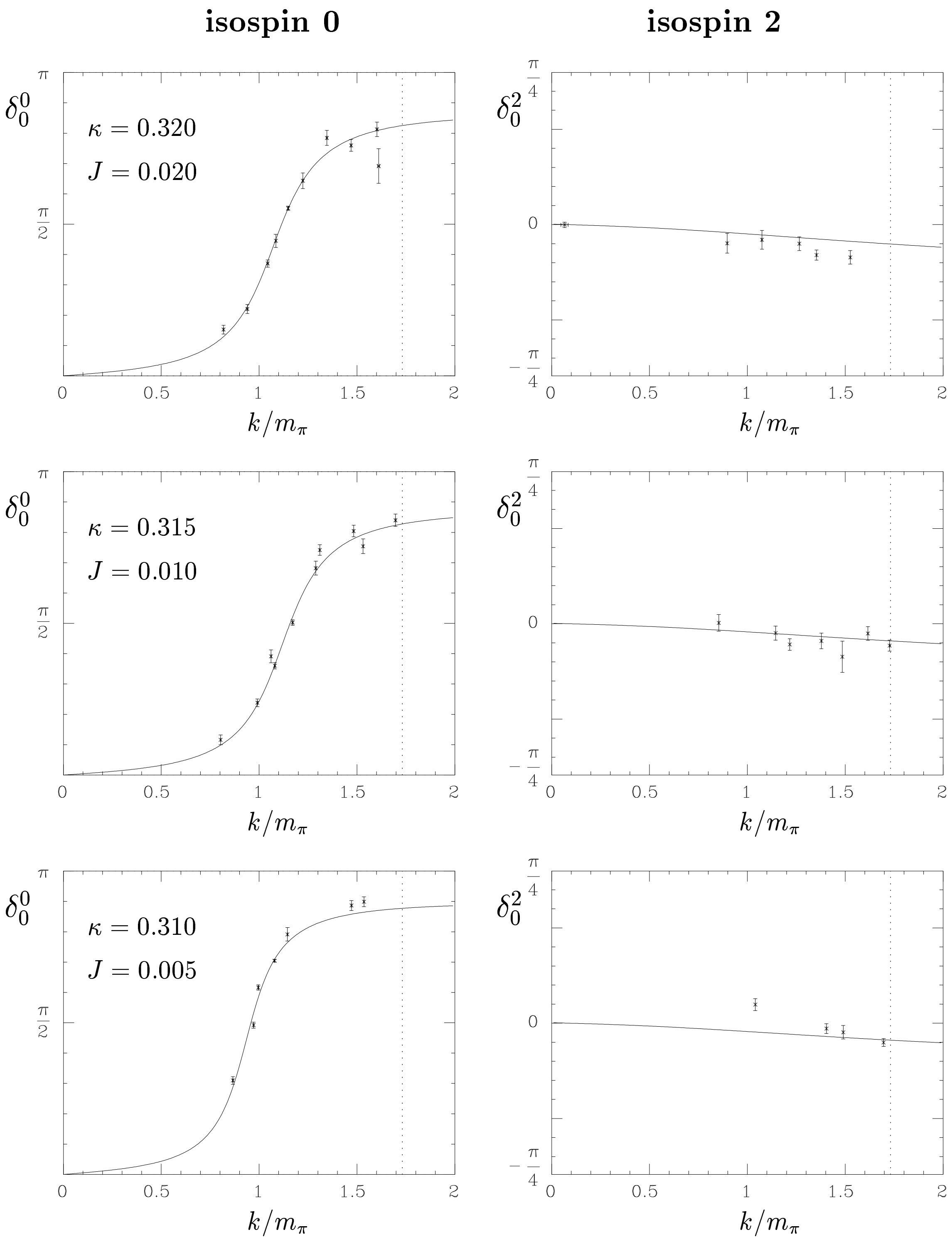

**Figure 17**: Scattering phase shifts for isospin 0 (l.h.s.) and for isospin 2 (r.h.s.). In the case of isospin 0 the data are very well explained by a fit according to the perturbative ansatz (3.6), and the values of the scattering phase shift in the isospin-2 channel are in good agreement with the perturbative calculations based on these results (solid lines).

# Scattering phases on finite lattices in the broken phase of the four-dimensional O(4) $\Phi^4$ theory*

Meinulf Göckeler[1,2], Hans A. Kastrup[1], Jörg Westphalen[1,2], Frank Zimmermann[1,2]

[1]*Institut für Theoretische Physik E, RWTH Aachen, D-52056 Aachen, Germany*
[2]*HLRZ c/o KFA Jülich, P.O.Box 1913, D-52425 Jülich, Germany*

**Abstract**

According to a proposal of Lüscher it is possible to determine elastic scattering phases in infinite volume from the energy spectrum of two-particle states in a periodic box. We demonstrate the applicability of this method in the broken phase of the four-dimensional $O(4)$ non-linear $\sigma$-model in a Monte Carlo study on finite lattices. This non-perturbative approach also permits the study of unstable particles, the $\sigma$-particle in our case. We observe the $\sigma$-resonance and extract its mass and width. In all scattering channels investigated the results are completely consistent with perturbative calculations.

* Supported by the Deutsche Forschungsgemeinschaft

# 1 Introduction

Most experiments in High Energy Physics are scattering experiments, the analysis of which leads to scattering phases, especially in the elastic channel. Therefore, it is a major challenge for theory to calculate these phases (or the corresponding scattering amplitudes). In the past, this has been done extensively by means of perturbative methods. However, if couplings become strong, non-perturbative methods are required. The most prominent example of a strongly coupled theory is QCD at low energies: Up to now it has been impossible to derive hadron-hadron scattering phases in an *ab initio* calculation from the fundamental theory (except for pion scattering lengths in the quenched approximation on the lattice [1–4]); in particular, the formation of resonances like, e.g., the $\rho$-meson, is still poorly understood.

The most promising non-perturbative approach in quantum field theory consists in putting the theory on a lattice and computing the functional integral by Monte Carlo simulations. Therefore, it is highly welcome to have a method that allows to extract elastic scattering phases from quantities which can be calculated in a Monte Carlo "experiment". Such a method has been devised by Lüscher [5,6]. It is based on a relation between the energy spectrum of two-particle states in a finite box with periodic boundary conditions and elastic scattering phase shifts defined in infinite volume. This relation is proved to hold for the scattering of two identical bosons below the inelastic threshold in any massive quantum field theory provided polarization effects are negligible. Since the spectrum of two-particle states in a finite volume can be computed numerically, we are thus in a position to determine phase shifts from Monte Carlo simulations. The fact that numerical simulations are necessarily performed on finite lattices is no disadvantage here, rather the finite volume is exploited to probe the system.

Scattering lengths have already been investigated [1–4,7–9] by means of the asymptotic large-volume behaviour of the lowest two-particle energy levels [10,11]. Furthermore, Lüscher's relation for two-dimensional models has been used to extract scattering phases in the O(3) nonlinear $\sigma$-model [12] and to study resonance scattering for two coupled Ising systems [13]. It has also been applied in three dimensions to meson-meson scattering in QED [14]. For previous attempts to investigate resonances on finite lattices see refs. [15,16].

In this paper we demonstrate the applicability of the procedure in the four-dimensional O(4)-symmetric $\Phi^4$-model. The model is studied in its broken phase, and the particles whose scattering is investigated are the three Goldstone bosons ("pions"). Since Lüscher's method only works in the absence of massless particles, we introduce an external source term which breaks the O(4) symmetry explicitly and hence makes the pions massive. However, the pion mass is kept small enough to allow for a resonance (the "$\sigma$-particle") in the appropriate channel and we are able to calculate the mass and the width of this resonance. Most likely our model is not strongly interacting in the scaling region, so we have perturbation theory to compare with. Indeed, we find impressive agreement. The discrepancy noted in our preliminary results [17] turned out to be caused by an inconsistent normalization [18].

The paper is organized as follows: Sect. 2 summarizes Lüscher's method. The perturbative results for the scattering phases are given in sect. 3. In sect. 4 we describe the details of our Monte Carlo simulation. After a discussion of the single-particle spectrum in sect. 5 we turn to the calculation of the two-particle energies (sect. 6). The results of these calculations are presented in sect. 7, followed by a brief discussion of the wave functions used in the construction of the two-particle operators (sect. 8). In sect. 9 we apply Lüscher's method and compute scattering phases from the measured two-particle energies. In particular, we obtain the mass and the width of the $\sigma$-resonance. Sect. 10 contains our conclusions.

# 2 Theoretical background

In this section we summarize the formulæ relating the energy spectrum of two-particle states in a finite periodic box and elastic scattering phases. We shall not give this relation in full generality, but restrict ourselves to the cases which are relevant for the applications we have in mind. For more details as well as for the proofs we refer to ref. [5].

Consider a system of two identical bosons of mass $m_\pi > 0$ with zero total momentum in a cubic box of size $L^3$. The two-particle states in the elastic region are characterized by centre-of-mass energies $W$ or relative momenta $k$ related by

$$W = 2\sqrt{m_\pi^2 + \vec{k}^2}\,, \qquad k = |\vec{k}\,|\,, \tag{2.1}$$

where $W$ lies in the range

$$2m_\pi < W < 4m_\pi \tag{2.2}$$

or, equivalently, with $k$ satisfying

$$0 < k < \sqrt{3}\,m_\pi\,. \tag{2.3}$$

Rotational invariance being broken by the cubic box, the two-particle states are classified according to the irreducible representations of the cubic group $O(3,\mathbb{Z})$. Below the inelastic threshold, the discrete energy spectrum corresponding to a particular representation is determined by the scattering phase shifts $\delta_l$ with angular momenta $l$ which are allowed by the symmetry of the states. The subspace of cubically invariant states (i.e. states transforming according to the trivial representation $A_1^+$) receives contributions from $l = 0, 4, 6, \ldots$ Similarly, the $T_1^-$-sector (vector representation) contains $l = 1, 3, \ldots$ We shall only be interested in the cases $l = 0$ and $l = 1$. Hence we have to study two-particle states transforming according to the representations $A_1^+$ and $T_1^-$ of the cubic group. Fortunately the contributions of the higher angular momenta in each sector will turn out to be negligible.

Assuming dominance of the lowest angular momentum in each symmetry sector, an energy value $W_\nu$ ($\nu = 0, 1, 2, \ldots$) belongs to the two-particle spectrum in a finite volume, if the corresponding momentum $k_\nu = \sqrt{(W_\nu/2)^2 - m_\pi^2}$ is a solution of

$$\delta_l(k_\nu) = -\phi\left(\frac{k_\nu L}{2\pi}\right) \mod \pi \tag{2.4}$$

for $l = 0$ in the case of the $A_1^+$-sector and for $l = 1$ in the case of the $T_1^-$-sector (see ref. [6]). The function $\phi(q)$ is defined as the continuous function satisfying

$$\tan(-\phi(q)) = \frac{q\pi^{3/2}}{\mathcal{Z}_{00}(1;q^2)}\,,\ \phi(0) = 0\,. \tag{2.5}$$

The zeta function $\mathcal{Z}_{00}(s;q^2)$ is given by (also for $q^2 < 0$)

$$\mathcal{Z}_{00}(s;q^2) = \frac{1}{\sqrt{4\pi}} \sum_{\vec{n}\in Z^3} \left(\vec{n}^2 - q^2\right)^{-s} \tag{2.6}$$

for Re $s$ sufficiently large and can be continued analytically to $s = 1$. The above statement holds only for $\left(\frac{kL}{2\pi}\right)^2 < 9$, which is fulfilled in our simulations. Note that eq.(2.4) has been derived in the continuum, hence it may be applied to lattice results only if scaling violations can be ignored. Furthermore, the derivation neglects finite volume polarization effects (exchange of particles "around the world"). Whether this is justified, can be checked by a careful study of the single-particle states.

In our Monte Carlo investigation we calculate the two-particle energy spectrum $W_\nu$ in a sector of definite cubic symmetry ($A_1^+$ or $T_1^-$) for a given size $L$. The scattering phase at the corresponding momentum can then be read off from eq.(2.4). In order to scan the momentum dependence of the phase shift we vary the spatial extent $L$ of the lattice.

Conversely, if $\delta_l(k)$ is considered to be known (e.g. from perturbation theory), eq. (2.4) allows us to compute the momenta $k_\nu$ and hence the two-particle energy spectrum for a given box size $L$. Fig. 1 illustrates the graphical solution of eq. (2.4) using the lowest order perturbative result for $\delta_0(k)$ (in the isospin-0 channel, see following section) as input: The momenta $k_\nu$ corresponding to the energy values $W_\nu$ are determined by the intersections of the solid curve ($\phi(q)$) with the dashed lines ($\delta_0(k) - \nu\pi$ for $\nu = 0, 1, 2, \ldots$). Note that in this example the lowest energy state ($\nu = 0$) is calculated from the analytical continuation of the scattering phase to imaginary values of $k$.

# 3 Perturbative results for the scattering phases

Most likely, the $\Phi^4$ model in four dimensions is trivial, and hence the renormalized coupling is bounded from above (for a recent review see ref. [19]). Indeed the bound turns out to be so low that the theory can be considered as weakly interacting in the whole scaling region (at least in the lattice regularization used in this work) [20]. Therefore perturbation theory should be applicable and lead to predictions for the scattering phases.

Considering the $O(N)$-symmetric $\Phi^4$-theory we write the action in the euclidean continuum as

$$S\{\vec{\phi}; j, m^2\} = \int d^4x \left[\frac{1}{2}\partial_\mu\phi^\alpha\partial_\mu\phi^\alpha + \frac{1}{2}m^2\phi^\alpha\phi^\alpha + \frac{g}{4!}\left(\phi^\alpha\phi^\alpha\right)^2 - j\phi^N\right] \tag{3.1}$$

with $\alpha = 1, \ldots, N$. Since we work in the broken phase, the mass parameter $m^2$ is negative. The external field, which breaks the symmetry down to $O(N-1)$, is taken to point in the $N$–direction and has the magnitude $j$.

The pions, whose scattering is to be investigated, are characterized by their momenta $\vec{k}$ and an "isospin" index $a = 1, \ldots, N-1$. The single-pion states transform according to the vector representation (isospin 1 for $N=4$) of the residual $O(N-1)$ symmetry. The product of two of these representations decomposes into three irreducible representations (with isospin $I = 0, 1, 2$, respectively, for $N=4$). The corresponding projectors $Q^I$ are given by

$$\begin{aligned}
Q^0_{a'b',ab} &= \frac{1}{N-1}\,\delta_{a'b'}\delta_{ab}\,, \\
Q^1_{a'b',ab} &= \frac{1}{2}\,(\,\delta_{a'a}\delta_{b'b} - \delta_{a'b}\delta_{b'a}\,) \\
Q^2_{a'b',ab} &= \frac{1}{2}\,(\,\delta_{a'a}\delta_{b'b} + \delta_{a'b}\delta_{b'a}\,) - \frac{1}{N-1}\delta_{a'b'}\delta_{ab}\,.
\end{aligned} \tag{3.2}$$

With their help, the amplitude $T$ for elastic two-pion scattering can be expressed in terms of the scattering amplitudes $T^I$ for fixed isospin $I$:

$$T = \sum_{I=0}^{2} Q_I\, T^I\,. \tag{3.3}$$

In the centre-of-mass system with total energy $W = 2\sqrt{m_\pi^2 + \vec{k}^2}$ the amplitude $T^I$ is a function of the scattering angle $\theta$ and the absolute value of the momentum $k = |\vec{k}|$. The partial wave decomposition reads:

$$T^I = \frac{16\pi W}{k} \sum_{l=0}^{\infty} (2l+1)\; P_l(\cos\vartheta)\; t_l^I(k)\,, \tag{3.4}$$

where $P_l$ denotes the Legendre polynomials. Note that due to Bose symmetry the partial wave amplitudes $t_l^I$ vanish, if $I+l$ is odd. In the elastic region $2m_\pi < W < 4m_\pi$ we can express $t_l^I(k)$ in terms of the real scattering phase $\delta_l^I(k)$:

$$t_l^I = \frac{1}{2i}\,(\,e^{2i\delta_l^I} - 1\,) \tag{3.5}$$

(provided the S-matrix is unitary).

We are interested only in the case where the $\sigma$-particle is unstable ($m_\sigma > 2m_\pi$) and appears as a resonance in the corresponding channel ($I = 0,\ l = 0$). The scattering amplitude $t_0^0$ becomes singular and the perturbative calculation requires a separation of the singular contribution $\delta^0_{0,s}$ to $\delta_0^0$ from the regular piece $\delta^0_{0,r}$ [21].

In lowest nontrivial order one obtains the following results [21]:

$$\delta_0^0 = \delta^0_{0,r} + \delta^0_{0,s} \tag{3.6}$$

$$\text{with} \qquad \tan\delta^0_{0,s} = g_R \frac{N-1}{48\pi}\,\frac{m_\sigma^2 - m_\pi^2}{m_\sigma^2 - W^2}\,\frac{k}{W}\,,$$

$$\text{and} \qquad \delta^0_{0,r} = \delta_0^2 - g_R \frac{N-1}{48\pi}\,\frac{k}{W}\,,$$

$$\delta_0^2 = \frac{g_R}{96\pi}\,\frac{m_\sigma^2 - m_\pi^2}{kW}\ln\Big(\,\frac{4k^2 + m_\sigma^2}{m_\sigma^2}\,\Big) - \frac{g_R}{24\pi}\,\frac{k}{W}\,, \tag{3.7}$$

$$\delta_1^1 = \frac{g_R}{96\pi}\frac{m_\sigma^2 - m_\pi^2}{kW}\Big(\,1 + \frac{m_\sigma^2}{2k^2}\,\Big)\ln\Big(\,\frac{4k^2 + m_\sigma^2}{m_\sigma^2}\,\Big) - \frac{g_R}{48\pi}\frac{m_\sigma^2 - m_\pi^2}{kW}\,. \tag{3.8}$$

Here $m_\pi$ and $m_\sigma$ denote the physical masses of the particles (the real part of the propagator pole in the case of the $\sigma$) and $g_R$ is the renormalized coupling constant. Comparison of $\delta_0^0$, $\delta_0^2$ with $\delta_l^0$, $\delta_l^2$ for $l \geq 4$ shows that the higher angular momenta are completely negligible in perturbation theory. The same statement holds in the $I=1$ channel: $\delta_l^1$ with $l \geq 3$ is much smaller than $\delta_1^1$.

The perturbatively calculated scattering phase $\delta_0^0$ may be continued analytically to purely imaginary values of $k$. In the $I=0$ channel this leads to scattering states which are obtained as solutions of eq.(2.4) for negative $q^2 = (kL/2\pi)^2$.

Fig. 2 shows an example of a perturbatively predicted two-particle energy spectrum. The dotted lines indicate the free energy spectrum

$$\frac{W}{m_\pi} = 2\sqrt{1 + \frac{\vec{k}^2}{m_\pi^2}}, \quad \vec{k} = \frac{2\pi}{L}\vec{n}, \quad \vec{n} \in \mathbb{Z}^3 . \tag{3.9}$$

# 4 The Monte Carlo simulation

On the lattice we parameterize the action as follows:

$$\begin{aligned} S\{\vec{\Phi}; J, \kappa\} &= \sum_x \Big\{ -\kappa \sum_{\mu=1}^{4} \Big( \Phi^\alpha(x)\Phi^\alpha(x+\hat{\mu}) + \Phi^\alpha(x)\Phi^\alpha(x-\hat{\mu}) \Big) \\ &\qquad + \lambda \Big( \Phi^\alpha(x)\Phi^\alpha(x) - 1 \Big)^2 + \Phi^\alpha(x)\Phi^\alpha(x) - J\Phi^N(x) \Big\} , \end{aligned} \tag{4.1}$$

Here we assume a hypercubic lattice whose spacing is put equal to 1. As usual, $\hat{\mu}$ denotes the unit vector pointing in $\mu$-direction. The relation between the lattice parameterization (4.1) and the continuum parameterization (3.1) of the action is given by the formulæ:

$$\begin{aligned} \phi^\alpha(x) &= \sqrt{2\kappa}\, \Phi^\alpha(x) , & j &= \frac{1}{\sqrt{2\kappa}}\, J , \\ -m^2 &= \frac{2\lambda - 1}{\kappa} + 8 , & g &= \frac{6\lambda}{\kappa^2} . \end{aligned} \tag{4.2}$$

In the actual simulation we used $N=4$ and $\lambda = \infty$ so that the scalar field is represented as a real four-component vector $\vec{\Phi}(x)$ of unit length: $\Phi^\alpha(x)\Phi^\alpha(x)=1$.

We shall work on lattices of size $L^3 \cdot T$ using a cluster algorithm. The simulation parameters $\kappa$, $J$, $L$, $T$ have to be chosen judiciously in order to extract a maximum of information from a given amount of computer time. As guideline we use the results on the scaling behaviour of our model obtained in ref. [22].

## 4.1 Choice of the simulation parameters

On the basis of the scaling laws (see ref. [22]) we plot in fig. 3 lines of constant mass ratios $m_\sigma/m_\pi$ in the $(\kappa, J)$-plane. In order to be sure that the $\sigma$-resonance lies in the elastic region $2m_\pi < W < 4m_\pi$, we want to choose the parameters $\kappa$ and $J$ such that $m_\sigma/m_\pi \approx 3$ (solid line in fig. 3). Furthermore, we have to suppress vacuum polarisation effects by making $m_\pi L$ sufficiently large. The dotted line in fig. 3 serves to illustrate this constraint: Above this line we have $m_\pi L \gtrsim 3$ for $L=16$. Finally we ended up with the choices $(\kappa,\ J)=(0.310,\ 0.005)$, $(0.315,\ 0.010)$ and $(0.320,\ 0.020)$ shown as black dots in fig. 3.

The spatial extent $L$ of the lattice should now be taken such that the corresponding momenta $k$ (see eq. (2.4)) cover the interesting region around the resonance where the phase shift is about $\pi/2$. For an estimate we put the resonance energy $W=m_\sigma$ leading to a relative momentum

$$k_\sigma = \sqrt{\frac{m_\sigma^2}{4} - m_\pi^2}\,. \tag{4.3}$$

Then the scattering phase shift $\delta_0^0$ takes the value $\pi/2$, if

$$L = \frac{2\pi q_{(0)}}{k_\sigma} \tag{4.4}$$

with $q_{(0)}$ such that $\phi(q_{(0)})=\pi/2 \pmod \pi$. According to eq.(2.5) these values $q_{(0)}$ are just the zeros of the zeta-function $\mathcal{Z}_{00}(1;q^2)$. Table 1 contains a list of the first of these zeros.

| index | $q_{(0)}^2$ |
|---|---|
| −1 | −0.095900719 |
| 0 | 0.472894247 |
| 1 | 1.441591313 |
| 2 | 2.627007612 |
| 3 | 3.536619947 |
| 4 | 4.251705973 |
| 5 | 5.537700774 |
| 6 | 7.196263202 |
| 7 | 8.287953655 |
| 8 | 9.534531427 |

**Table 1**: Zeros of the zeta-function $\mathcal{Z}_{00}(1;q^2)$.

| | $(\kappa,\ J)$ | $\tilde{m}_\sigma$ | $m_\pi$ | $\tilde{m}_\sigma/m_\pi$ | $L$ |
|---|---|---|---|---|---|
| **1** | (0.310, 0.005) | 0.5383(11) | 0.1844(9) | 2.88 | 22, 38 |
| **2** | (0.315, 0.010) | 0.7202(13) | 0.2309(8) | 3.13 | 16, 27, 37 |
| **3** | (0.320, 0.020) | 0.9059(16) | 0.2977(10) | 3.07 | 13, 22, 30, 35, 38 |

**Table 2**: Numerically realizable lattice extensions $L$ for which the two-particle energy spectrum has an energy level in the resonance region.

| | $(\kappa,\, J)$ | LATTICE SIZES $L^3 \cdot T$ | | | | | |
|---|---|---|---|---|---|---|---|
| **1** | (0.310, 0.005) | | $16^3 \cdot 32$, | $20^3 \cdot 32$, | $24^3 \cdot 32$, | | $32^3 \cdot 40$ |
| **2** | (0.315, 0.010) | | $16^3 \cdot 32$, | $20^3 \cdot 32$, | $24^3 \cdot 32$, | $24^3 \cdot 60$, | $32^3 \cdot 40$ |
| **3** | (0.320, 0.020) | $12^3 \cdot 32$, | $16^3 \cdot 32$, | $20^3 \cdot 32$, | $24^3 \cdot 32$ | | |

**Table 3**: Actually realized simulation parameters $(\kappa,\, J)$ and lattice sizes $L^3 \cdot T$.

From test runs or from the scaling laws [22] we get estimates for $m_\pi$, $m_\sigma$ and hence $k_\sigma$. (Actually, we took for $m_\sigma$ the mass $\tilde{m}_\sigma$ extracted from a fit of the $\sigma$-propagator in momentum space.) With the help of table 1, eq.(4.4) then leads to a list of $L$ values for which a two-particle energy level close to the resonance energy is to be expected (see table 2). In particular, this list gives us the minimal useful $L$-values. Also taking into account that the computing resources limit $L$ from above, we adjusted the simulation parameters such that at least four different lattice sizes could be used. Finally choosing $T > L$ (to allow for a reliable determination of energy levels) we arrive at the simulation parameters shown in table 3.

## 4.2 Technical details

Our configurations are generated by means of the cluster algorithm [24,23] generalized for actions with finite external source: Once the multi-cluster structure $\{c_j\}$ is determined, in the standard multi-cluster method (for $J=0$) the spins of every second cluster are reflected with respect to the hyperplane perpendicular to a randomly chosen global $O(4)$-direction $\vec{r}$. Generalized to a finite external source $J$ the clusters $c_j$ are flipped with probability

$$w(\vec{\Phi}, c_j) = \Big( 1 + \exp\Big( 2Jr_4 \sum_{x \in c_j} (\vec{r} \cdot \vec{\Phi}(x)) \Big) \Big)^{-1} . \tag{4.5}$$

depending on the field configuration within the cluster $c_j$.

The simulations are performed on lattices of size $L^3 \cdot T$. On the HLRZ-CRAY YMP8/832 we can only run lattices of sizes $12^3 \cdot 32$, $16^3 \cdot 32$ and $20^3 \cdot 32$. The larger lattices $24^3 \cdot 32$, $24^3 \cdot 60$ and $32^3 \cdot 40$ are simulated on the *Landesvektorrechner* Fujitsu/Siemens-Nixdorf SNI S600/20 in Aachen. The speed of our simulation program is about $180-200$ MFlops on the CRAY and $500-650$ MFlops on the Fujitsu. We got a mean computing time of about $6.0 \cdot 10^{-6}$ seconds per lattice site and iteration on the CRAY and $1.8 \cdot 10^{-6}$ seconds on the SNI S600/20.

As a rule we performed 150 000 to 330 000 iterations per simulation point $(\kappa,\, J)$ and lattice size $L^3 \cdot T$, measuring after every fourth iteration. Averages over blocks of 1024 measurements are stored on disk. Of course, a later increase of the block length is possible. In order to generate the random number sequences we used the Kirkpatrick-Stoll-Greenwood (shift register) random number generator which vectorizes on both machines [25].

| | Simulation points | | |
|---|---|---|---|
| | **1** | **2** | **3** |
| $\kappa$ | 0.310 | 0.315 | 0.320 |
| $J$ | 0.005 | 0.010 | 0.020 |
| $m_\pi$ | 0.1870(6) | 0.2300(6) | 0.2947(6) |
| $\tilde{m}_\sigma$ | 0.5383(11) | 0.7202(13) | 0.9059(16) |
| $\Sigma$ | 0.18455(3) | 0.23574(2) | 0.28096(2) |
| $Z$ | 0.9745(5) | 0.9723(5) | 0.9715(5) |
| $Z_\sigma$ | 0.9323(10) | 0.9130(12) | 0.8966(13) |
| $\tilde{g}_R$ | 21.96(13) | 24.43(12) | 27.03(14) |
| $c_p$ | −0.0614(4) | −0.0634(4) | −0.0602(2) |

**Table 4**: Infinite volume results used in order to compute the perturbative predictions of the two-particle energy spectrum.

# 5 Pion mass and single-particle spectrum

Defining the particle mass $m_\pi$ by the single-particle energy with spatial momentum $\vec{p}=0$ in infinite volume we expect the relativistic energy momentum relation

$$E(\vec{p}) = \sqrt{m_\pi^2 + \vec{p}^{\,2}} \tag{5.1}$$

also to be valid on a finite lattice provided $m_\pi L \gg 1$ and $|\vec{p}| \ll 1$. The first inequality insures the suppression of finite volume effects [26] and the second one is related to the lattice effects, which for our choice of lattice action decrease with the square of the lattice spacing.

In order to control both polarization effects and scaling violations, we first examine the single-particle energies extracted from the exponential decay of the propagator

$$\left\langle \sum_{a=1}^{3} \sum_{\vec{n}\in\mathbb{Z}_L^3} \delta_{j,\vec{n}^2}\, \tilde{\Phi}^a(-\vec{n},t)\, \tilde{\Phi}^a(\vec{n},0) \right\rangle \sim e^{-E(\vec{p})t} \tag{5.2}$$

with $\vec{p}^{\,2} = (2\pi/L)^2\, \vec{n}^{\,2} = (2\pi/L)^2\, j$. Here $\tilde{\Phi}^a(\vec{n},t)$ is the spatial Fourier transform of the pion field $\Phi^a(\vec{x},t)$. For a general lattice function $F(\vec{x})$ it is defined by:

$$\tilde{F}(\vec{n}) = \frac{1}{L^3} \sum_{\vec{x}\in\mathbb{Z}_L^3} e^{i\,2\pi\,\vec{x}\cdot\vec{n}/L} F(\vec{x})\,. \tag{5.3}$$

| $j$ | $\vec{n}$ | $\lvert\vec{p}\rvert/m_\pi$ | $E(\vec{p})/\sqrt{m_\pi^2+\vec{p}^2}$ | $E(\vec{p})/E_{lat}(\vec{p})$ |
|---|---|---|---|---|
| 1 | (1,0,0) | 0.8539 | 0.9932(12) | 0.9976(12) |
| 2 | (1,1,0) | 1.2076 | 0.9924(8) | 0.9987(8) |
| 3 | (1,1,1) | 1.4790 | 0.9893(8) | 0.9973(8) |
| 4 | (2,0,0) | 1.7078 | 0.9885(6) | 1.0016(6) |
| 5 | (2,1,0) | 1.9093 | 0.9859(4) | 1.0000(4) |

**Table 5**: Single-pion energy levels and their deviation from the relativistic continuum and lattice dispersion relation as function of $|\vec{p}|$ for the simulation parameters ($\kappa = 0.315$, $J = 0.01$, $L = 32$). In the second column we give the lattice momenta $\vec{n}$ with $\vec{n}^2 = j$ for which the energy is calculated. These momenta are used to compute $\vec{\hat{p}}$ in eq. (5.5).

The numerical data of the single-particle spectrum are well approximated by a fit of the form

$$E(\vec{p}) \approx \sqrt{m_\pi^2 + \vec{p}^2}\;\left(1 + c_p\,\vec{p}^2\right) \tag{5.4}$$

(see fig. 4). Alternatively the data can be compared with the lattice dispersion relation

$$E(\vec{p}) \approx E_{lat}(\vec{p}) = 2\,\mathrm{asinh}\left(\frac{1}{2}\sqrt{m_\pi^2 + \vec{\hat{p}}^2}\right)\;,\quad \hat{p}_i = 2\sin\frac{p_i}{2}\,. \tag{5.5}$$

A fit of the form (5.4) or (5.5) for fixed lattice extent $L$ yields $m_\pi(L)$ (and $c_p(L)$). The $L$-dependence of $c_p(L)$ turns out to be so weak that we simply average over the results for different values of $L$ to obtain our final estimate for $c_p$ in the infinite volume limit (see table 4). The pion mass, on the other hand, is extrapolated to infinite volume assuming a decay of the polarisation effects proportional to $\exp(-m_\pi L)/(m_\pi L)^2$. Both fits lead to consistent results for the pion mass. Table 4 contains the values obtained by means of (5.5). Determining $m_\pi$ by a fit to the single-particle energies we gain the advantage that all measured values of $E(\vec{p})$ contribute to the calculation.

Fig. 4 compares the measured single-pion energies $E(\vec{p})$ at simulation point 2 ($\kappa = 0.315$, $J = 0.01$) with the lattice dispersion relation (5.5) and the continuum dispersion relation (some examples are listed in table 5). In both cases the infinite volume mass is used. Hence the plot demonstrates not only the quality of the fits (5.4) and (5.5) but also the smallness of the finite-size effects.

Alternatively, the pion mass $m_\pi$ can be measured by a fit to the inverse propagator in four-dimensional momentum space:

$$[G_{aa}(p,-p)]^{-1} = Z^{-1}\left\{\,m_\pi^2 + \hat{p}^2\,\right\} \qquad a = 1,2,3\,. \tag{5.6}$$

The excellent quality of that fit is shown in fig. 5. After extrapolating to infinite volume we obtain within 1% deviation the same $m_\pi$-values as with the determination from the single-particle spectrum. We also get the pion wave function renormalization constant $Z$.

For the perturbative calculation we need not only $m_\pi$ but also the $\sigma$-mass $m_\sigma$ and the renormalized coupling $g_R$. The latter is defined by

$$g_R = 3\,Z\,\frac{m_\sigma^2 - m_\pi^2}{\Sigma^2}\,, \tag{5.7}$$

where $\Sigma$ is the field expectation value in infinite volume. We extract $\Sigma$ from the volume dependence of the expectation value of the magnetisation parallel to the external source as outlined in ref. [22]. The $\sigma$-mass will later be determined from the analysis of the scattering phase shift $\delta_0^0$. However, a first estimate $\tilde{m}_\sigma$ can be obtained by a fit to the $\sigma$-propagator in momentum space (see fig. 5). A volume dependence of $\tilde{m}_\sigma$ is not observed. The results for $\tilde{m}_\sigma$ as well as the ensuing estimates $\tilde{g}_R$ for the renormalized coupling are given in table 4, which also contains the wave function renormalization constant $Z_\sigma$ extracted from the $\sigma$-propagator.

# 6 How to calculate the two-particle energy spectrum

Following ref. [12] we extract the two-particle energy spectrum from a matrix of correlation functions of suitable operators, which couple to the two-particle states (with the appropriate quantum numbers). The states and operators are classified according to O(3)-isospin taking the values 0, 1 and 2 (see eq. (3.2)) and irreducible representations of the cubic group $O(3,\mathbb{Z})$ ($A_1^+$ or $T_1^-$). By definition the two-particle states are energy eigenstates and in general not identical to the states generated by the application of the operators to the vacuum. The spectral decomposition of the correlation function matrix will finally enable us to calculate the two-particle energy spectrum, if the number of operators exceeds the number of energy eigenstates below the inelastic threshold in each symmetry sector.

## 6.1 Two-particle operators

Operators for two pions with total momentum zero are defined by the following double sum over the spatial part of the pion fields [12]

$$\mathcal{O}_i^{ab}(t) := \frac{1}{L^6} \sum_{\vec{x},\vec{y}\in\mathbb{Z}_L^3} f_i(\vec{x}-\vec{y})\ \Phi^a(\vec{x},t)\ \Phi^b(\vec{y},t)\,. \tag{6.1}$$

with suitable wave functions $f_i(\vec{x}-\vec{y})$ $(i = 1,2,\ldots)$. In order to save computer time we perform the actual calculations on the momentum lattice writing

$$\mathcal{O}_i^{ab}(t) = \sum_{\vec{n}\in\mathbb{Z}_L^3} \tilde{f}_i(\vec{n})\ \tilde{\Phi}^a(-\vec{n},t)\ \tilde{\Phi}^b(\vec{n},t) \tag{6.2}$$

with the spatial Fourier transform defined in eq. (5.3). Since the scalar fields $\Phi^a(\vec{x},t)$ are real, we have $\overline{\tilde{\Phi}^a(\vec{n},t)} = \tilde{\Phi}^a(-\vec{n},t)$, where a bar over a quanitity means complex conjugation. This implies: $\overline{\mathcal{O}_i^{ab}(t)} = \mathcal{O}_i^{ba}(t)$ for real wave functions $\tilde{f}_i(\vec{n})$ in momentum space[2].

[2] Or $\overline{f_i(\vec{x})} = f_i(-\vec{x})$ in coordinate space.

The Fourier transformation $\tilde{\Phi}^a(\vec{n},t)$ of the field is calculated by a three-dimensional *Fast Fourier Transformation* (FFT) which has excellent vectorization properties [27].

For isospin 0 and 2 we compare two different kinds of cubically invariant wave functions. First we take plane waves (as refs. [12, 13] in their two-dimensional models):

$$\tilde{f}_i(\vec{n}) = \frac{\delta_{j,\vec{n}^2}}{\sum\limits_{\vec{m}\in\mathbb{Z}_L^3} \delta_{j,\vec{m}^2}} \,. \tag{6.3}$$

The index $i = 0, 1, 2, \ldots$ counts the integers $j$ for which there is a three-dimensional integer vector $\vec{n}$ with $\vec{n}^2 = j$.[3]

Our second choice is motivated by the periodic singular solutions of the Helmholtz-equation (see ref. [5]):

$$\tilde{f}_i(\vec{n}) = \frac{(\vec{n}^2 - q_i^2)^{-1}}{\sum\limits_{\vec{n}\in\mathbb{Z}_L^3} (\vec{n}^2 - q_i^2)^{-1}} \,, \quad i = 0, 1, 2, \ldots, \tag{6.4}$$

where $q_i = (L/2\pi)\sqrt{(W_i/2)^2 - m_\pi^2}$ using as input the plane wave energy spectrum $W_i$ of some test runs. Another way to calculate estimates $q_i$ uses the results of table 4 together with perturbation theory: By computing the scattering phase shifts (3.6) and (3.7) and solving eq. (2.4) for $k_i$ one could estimate some values $q_i = k_i L/2\pi$. The wave functions (6.4) could have an enhanced projection especially in the isospin-0 channel at energies close to the resonance mass.

Later we will show that the calculated two-particle energy spectrum does not depend significantly on the choice of the wave function. Hence both sets of operators allow a reliable determination of the energy values below the inelastic threshold.

In the case of isospin 1 the analog of the cubically invariant wave functions (6.3) and (6.4) of the $A_1^+$ sector are the vector-like wave functions $\sim \vec{n}\,\delta_{j,\vec{n}^2}$ or $\sim \vec{n}/(\vec{n}^2 - q_i^2)$ of the $T_1^-$ sector. For our calculations we have adopted the first possibility.

## 6.2 Correlation function matrix

For each isospin we examine the (connected) correlation function matrix of the two-particle operators $\mathcal{O}_i^{ab}(t)$:

$$\begin{aligned} C_{ij}^I(t) &:= Q^I_{a'b',ab} \left\langle \overline{\mathcal{O}_i^{a'b'}(t)}\, \mathcal{O}_j^{ab}(0) \right\rangle_c \\ &= Q^I_{a'b',ab} \left( \left\langle \overline{\mathcal{O}_i^{a'b'}(t)}\, \mathcal{O}_j^{ab}(0) \right\rangle - \left\langle \overline{\mathcal{O}_i^{a'b'}(t)} \right\rangle \left\langle \mathcal{O}_j^{ab}(0) \right\rangle \right) . \end{aligned} \tag{6.5}$$

Since we expect to find the $\sigma$-resonance in the isospin-0 channel, we first restrict ourselves to that case. The corresponding two-pion operators $O_i(t)$ are given by the trace of the operators (6.2) in the space of the internal O(3)-symmetry (see eq. (3.2)):

$$O_i(t) = \frac{1}{\sqrt{3}} \sum_{a=1}^{3} \mathcal{O}_i^{aa}(t) \,. \tag{6.6}$$

[3] $j \neq i$ only for $j \geq 7$.

We also have to take into account the $\sigma$ field at zero momentum, since it has the same quantum numbers as the operators $O_i$ and is expected to create a state with energy below the inelastic threshold:

$$O_\sigma(t) = \tilde{\Phi}^4(\vec{0},t) = \frac{1}{L^3} \sum_{\vec{x}\in\mathbb{Z}_L^3} \Phi^4(\vec{x},t)\,. \tag{6.7}$$

With the help of eq. (3.2) we obtain from the definition of the correlation function matrix (6.5) the following expressions for the different isospin channels:

$$\begin{aligned} C^0_{ij}(t) &= \left\langle\, O_i(t)\; O_j(0) \,\right\rangle_c && i,j=\sigma,0,1,2,\ldots, \\ C^1_{ij}(t) &= \left\langle\, \mathrm{Im}\,\mathcal{O}^{ab}_i(t)\; \mathrm{Im}\,\mathcal{O}^{ab}_j(0) \,\right\rangle_c && i,j=1,2,\ldots, \\ C^2_{ij}(t) &= \left\langle\, \mathrm{Re}\,\mathcal{O}^{ab}_i(t)\; \mathrm{Re}\,\mathcal{O}^{ab}_j(0) \,\right\rangle_c - C^0_{ij}(t) && i,j=0,1,2,\ldots, \end{aligned} \tag{6.8}$$

where sums over repeated indices $a,b$ are implied. Disconnected contributions are of course only expected for isospin 0.

Instead of working with the connected correlation functions according to their definition we eliminate the disconnected contributions by measuring correlation function matrices of the following type[4] [12]:

$$\mathcal{C}_{ij}(t) = \left\langle\, (\,\mathcal{O}_i(t) - \mathcal{O}_i(t+1)\,)\;\; \mathcal{O}_j(0) \,\right\rangle\;. \tag{6.9}$$

From the transfer matrix formalism one gets the spectral decomposition of these correlation function matrices:

$$\mathcal{C}_{ij}(t) = \sum_{\nu=0}^{\infty} \overline{v^\nu_i}\; v^\nu_j\; e^{-tW_\nu} \quad \text{with} \quad v^\nu_j = \sqrt{1-e^{-W_\nu}}\,\langle\, \nu\, |\mathcal{O}_j(0)|\, \Omega\,\rangle\;. \tag{6.10}$$

The energy eigenvalues $W_\nu$ are assumed to be non-degenerate (in the symmetry sector under consideration) and are ordered such that $W_\nu < W_{\nu+1}$. The amplitudes $v^\nu_j$ in the spectral decomposition (6.10) are proportional to the projections of the states generated by the two-particle operator out of the vacuum $|\,\Omega\rangle$ onto the energy eigenstates $|\,\nu\,\rangle$.

Below the inelastic threshold $4m_\pi$ there is only a finite number $\bar{\imath}$ of energy eigenstates $W_\nu$. Furthermore, we can consider the correlation function matrix (6.10) only for a finite number $\bar{r}$ of indices $i,j$. This restriction is not only due to the finiteness of computing resources, but is also enforced by the fact that the number of linearly independent two-particle operators is finite on a finite lattice. In order to guarantee that the $\bar{r}$-component vectors $v^\nu$, $\nu = 0,1,2,\ldots,\bar{\imath}-1$, are linearly independent $\bar{r}$ has to be larger than the number $\bar{\imath}$ of (expected) states below the inelastic threshold.

For large time distances $t$ the eigenvalues of the correlation function matrix $\mathcal{C}(t)$ are proportional to $e^{-tW_\nu}$. Hence one could determine the energy eigenvalues from the eigenvalues of $\mathcal{C}(t)$ for $t\to\infty$. However, since the statistical errors of the matrix $\mathcal{C}(t)$ are increasing with $t$, we use the following method to extract the energy spectrum. It allows a reliable determination of the energy levels $W_\nu$ already for smaller values of $t$ [12]. Given a (small) reference time $t_0$ we consider the generalized eigenvalue problem

$$\mathcal{C}(t)\,\mathbf{w}^\nu = \lambda_\nu(t,t_0)\;\mathcal{C}(t_0)\,\mathbf{w}^\nu\,. \tag{6.11}$$

[4] Isospin indices are not shown explicitly in the following.

The eigenvalues $\lambda_\nu(t,t_0)$ are given by

$$\lambda_\nu(t,t_0) = e^{-(t-t_0)\,W_\nu} \tag{6.12}$$

up to (negligible) corrections of order $e^{-(t-t_0)\,W_{\bar{r}}}$. In order to analyse the generalized eigenvalue problem (6.11) we study the matrix

$$\mathcal{D}(t,t_0) = \mathcal{C}^{-\frac{1}{2}}(t_0)\;\mathcal{C}(t)\;\mathcal{C}^{-\frac{1}{2}}(t_0) \tag{6.13}$$

having the same eigenvalues $\lambda_\nu(t,t_0)$ but eigenvectors $\mathbf{u}^\nu = \mathcal{C}^{\frac{1}{2}}(t_0)\mathbf{w}^\nu$. To calculate the spectral amplitudes $v_j^\nu$ we solve the equations

$$\sum_i v_i^\nu\; w_i^\mu = \delta^{\mu\nu}\, e^{W_\nu t_0/2} \tag{6.14}$$

by matrix inversion. Like (6.12) these relations suffer from exponentially suppressed corrections.

The projection properties of the chosen operators onto the energy eigenstates are characterized by the normalized amplitudes

$$c_j^\nu = \frac{\tilde{v}_j^\nu}{\sqrt{\sum_\mu |\tilde{v}_j^\mu|^2}}\,, \qquad \tilde{v}_j^\nu = \frac{1}{\sqrt{1-e^{-W_\nu}}} v_j^\nu\,. \tag{6.15}$$

They do not depend on the normalization of the operators $\mathcal{O}_j$.

For the calculation of the correlation functions we have also applied the two-cluster method [12]. But eventually we did not adopt this method because of two reasons: It takes too much computer time for other wave functions than plane waves (6.3) and for the most interesting isospin-0 case we do not expect (and actually did not find) any reduced variance [12]. However, we tested this method and obtained agreement with the standard measurements within statistical errors.

## 6.3 Various methods of analysis

In order to control systematic errors and to improve the quality of the analysis we examined different methods to determine the two-particle energy spectrum (for a summary see fig. 6). They all lead essentially to the same results as is demonstrated in table 6 for a typical case.

As described above, we diagonalize the matrix $\mathcal{D}(t,t_0) = \mathcal{C}^{-\frac{1}{2}}(t_0)\mathcal{C}(t)\mathcal{C}^{-\frac{1}{2}}(t_0)$ for some small value of the reference time $t_0$. This reference time has to be kept small in order to guarantee the numerical stability of the determination of the eigenvalues.

To determine the energy spectrum $W_\nu$ we fit the eigenvalues with the exponential ansatz[5]

$$\lambda_\nu(t,t_0) = e^{-(t-t_0)\,W_\nu}\,. \tag{6.16}$$

[5]Using a fit with the function $\cosh\left((t-t_0-T/2)\,W_\nu\right)$ does not affect the results.

The data are combined into blocks, we take the block mean values and calculate the total mean values and errors using the Jackknife method. The data are written out by the simulation program with a standard block length $l_{blk}$ of 1024 measurements. In order to get a correct estimate of the errors we increase the block length $l_{blk}$ until the errors stabilize. Already with $l_{blk} = 1024$ the analysis seems to be stable, and we decided to put $l_{blk} = 2048$ in our final results.

There are (at least) three different routes to take the mean of the data: We can combine into blocks the correlation function matrices $\mathcal{D}(t, t_0)$, their eigenvalues $\lambda_\nu(t, t_0)$ or the energy levels $W_\nu$ from the exponential fit. These different strategies summarized in fig. 6 are denoted by ①, ②, and ③. A change of strategy only leads to negligible fluctuations of the energy levels (see table 6). Since the error analysis in strategy ③ seems to be most reliable, we use this method for calculating our final results.

Another free parameter of the analysis is the dimension $\bar{r}$ of the correlation function matrix $\mathcal{D}(t, t_0)$, which has to be diagonalized. As long as $\bar{r}$ is larger than the number $\bar{\imath}$ of energy levels expected below the inelastic threshold we do not see any significant dependence of the energy spectrum on $\bar{r}$.

In order to check the dependence on the temporal lattice extent $T$, we have performed runs on a $24^3 \cdot 60$ lattice. The results do not differ significantly from those obtained on a $24^3 \cdot 32$ lattice.

From the energy eigenvalues $W_\nu$ we have to calculate the corresponding momenta $k_\nu$. Neglecting lattice effects one could use the continuum dispersion relation: $(W_\nu/2)^2 = m_\pi^2 + \vec{k}_\nu^2$. Alternatively, as for the single-particle spectrum, we can try to suppress the dependence on the lattice constant by using the lattice dispersion relation. The corresponding relation for the two-particle levels reads

$$\left( 2 \sinh \frac{(W_\nu/2)}{2} \right)^2 = m_\pi^2 + \vec{\hat{k}}_\nu^{\,2} . \tag{6.17}$$

Since the absolute values of the momenta $k_\nu$ are smaller than the energy levels $W_\nu$ below the inelastic threshold (especially for the lowest level with $k_0 \approx 0$) we can assume that the dominating part of the lattice effects is removed by replacing

$$\frac{W_\nu}{2} \to 2 \sinh \frac{(W_\nu/2)}{2} . \tag{6.18}$$

As expected the lattice effects grow with increasing distance from the critical point $(\kappa = \kappa_c = 0.30423(1), J = 0)$. For the lowest energy levels they remain smaller than one standard deviation. For the higher energy levels below the inelastic threshold the lattice effects are at most twice as large. As we will see in sect. 9 at the simulation point $(\kappa = 0.320,\ J = 0.02)$ with our largest pion mass (and therefore the highest energy values) this will give a visible difference in the calculated scattering phase shifts. Compared with the lattice effects the influence of the other uncertainties mentioned in this section is negligible (see table 6).

Due to the loss of rotational invariance on the lattice it is not completely unique how to calculate $k_\nu$ on the basis of eq. (6.17). We circumvent this problem by using the direction of the momentum corresponding to the nearest free energy level $j \approx (L/2\pi)^2 \vec{k}^2$ listed in table 5. The associated ambiguity might lead to a systematic error which is of the same size as the statistical error.

## 6.4 A self-adjusting exponential fit

We have developed a method to determine the upper end $t_{max}$ of the range of the exponential fit (6.16) automatically. The eigenvalues $\lambda_\nu(t,t_0)$ are sorted such that the corresponding eigenvectors $\mathbf{u}^\nu(t)$ of the correlation function matrix $\mathcal{D}(t,t_0)$ at successive values of $t$ are as parallel as possible: Ideally the eigenvector $\mathbf{u}^\nu$ corresponding to an energy level $W_\nu$ should be time-independent (see sect. 6.2). To get the optimal assignment of (almost parallel) eigenvectors is a standard problem of linear optimization. It can be solved starting from the matrix $A_{\nu\mu} = (\mathbf{u}^\nu(t), \mathbf{u}^\mu(t{+}1))$ of the scalar products of the eigenvectors on successive time slices:

$$\max \sum_{\nu,\mu} x_{\mu\nu}\, A_{\mu\nu} \tag{6.19}$$
$$\text{with} \qquad \sum_{\nu} x_{\mu\nu} = \sum_{\mu} x_{\mu\nu} = 1 \quad \text{and} \quad x_{\mu\nu} \in \{0,1\}\ .$$

We have approximated the exact solution by the following procedure: We determine the largest matrix element $A_{\bar\nu\bar\mu}$ and assign $\mathbf{u}^{\bar\nu}(t)$ to $\mathbf{u}^{\bar\mu}(t{+}1)$. After deleting the corresponding row and column we continue in this way with the truncated matrix.

The sorting of the eigenvalues according to the collinearity of their eigenvectors should be equivalent to a sorting with respect to their size, provided the eigenvalues at $t_0{+}1$ are arranged in accordance with this criterion. But due to statistical fluctuations at some time slice $t = t_{max}{+}1$ this fails, thus determining the upper end $t_{max}$ of the fit range (for an example see fig. 7).

## 6.5 Final choice of the method

Table 6 shows the weak dependence of the spectrum on the choice of variations displayed in fig. 6. The variations are done independently for each degree of freedom starting at a standard choice. From table 6 we learn to keep the reference time $t_0$ small and to guarantee a complete set of linear independent eigenvectors corresponding to the energy levels in the elastic region ($\bar r > \bar\imath$).

Among all the possibilities we selected the following list of parameters, which are labeled by a • in table 6:

- reference time: $t_0 = 0$
- block length: 2048 (with 18 − 40 blocks we obtain enough statistics)
- choice of the strategy ③ (taking the mean of the energy values)
- dimension of $\mathcal{D}(t,t_0)$: at least $\bar r = \bar\imath + 1$
- temporal lattice extent $T = 32$ and 40, respectively ($24^3 \cdot 60$ lattice only with $\kappa = 0.315$ and $J = 0.01$)
- "Lüscher's" two-particle wave function (6.4) with isospin 0 and plane waves (6.3) with isospin 1 and 2

For comparison we also illustrate the systematic shift of the energy values due to the replacement (6.18). The following section summarizes the results of our analysis of the two-particle energy spectrum obtained along these lines.

# 7 Numerical results for the two-particle energy spectrum

The correlation function matrices are analysed according to the criteria mentioned above. The results are shown in tables 7 to 10 without corrections for lattice effects. These final results are the starting point for the calculation of the scattering phase shifts and the determination of the resonance parameters $m_\sigma$ and $\Gamma_\sigma$ (and the coupling $g_R$, respectively) as discussed in the following section. For comparison we also show the perturbatively calculated energy spectrum using as input parameters the results of sect. 9 (see table 11) instead of the estimates in table 4.

Fig. 10 shows all data of the two-particle energy spectra for isospin 0 and 2 at a glance, whereas fig. 8 and 9 give an enlarged overview at the simulation point ($\kappa = 0.315$, $J = 0.010$):

- Crosses symbolize the measured energy values below the inelastic threshold $4m_\pi$. The errors are smaller than the symbols and lattice effects are taken into account by means of the replacement (6.18).
- The solid lines passing through our data represent the energy spectrum calculated perturbatively on the basis of the results of sect. 9. We observe good agreement with perturbation theory.
- The perturbative spectrum computed from the estimates of table 4 is indicated by the dashed curves in fig. 8

For isospin 1 the picture would look similar to the isospin 2 case except that there is no energy level around $2m_\pi$.

In the **isospin-0 resonance channel** we see the expected trend (cf. fig. 2): The first and second level are coming close to each other displaying the so-called "*avoided level crossing*" such that we get an impression where the resonance plateau might be (the dotted line at about $W \approx 3m_\pi$).

In fig. 8 we also compare our results with perturbative calculations based on the estimates of table 4. The reason why the corresponding dashed lines do not fit to the data is related to the relatively large width of the resonance (see sect.9): the naive fit to the propagator in momentum space does not yield the correct value for the resonance mass. However, there is very nice agreement between perturbation theory and our data, if we take the resonance parameters $m_\sigma$ and $\Gamma_\sigma$ ($g_R$) determined in sect. 9.

In the **isospin-1** and the **isospin-2 channel** the numerically calculated two-particle energy spectrum agrees very well with the perturbative prediction, which is close to the free energy spectrum and depends only weakly on $m_\sigma$ (see e.g. fig. 9 for isospin 2). As shown in fig. 10 we obtain a similar behaviour at all simulation points.

| $\kappa = 0.315$<br>$J = 0.010$ | **two-particle energy spectrum**<br>ISOSPIN 0 | | | | | | |
|---|---|---|---|---|---|---|---|
| | $W_0$ | $W_1$ | $W_2$ | $W_3$ | $W_4$ | $W_5$ | $W_6$ | |
| $t_0 = 0$ | 0.456(2) | 0.644(2) | 0.747(3) | 0.902(4) | 1.22(2) | | | • |
| $t_0 = 1$ | 0.452(4) | 0.64(1) | 0.75(1) | 0.86(2) | 1.25(7) | | | |
| $t_0 = 2$ | 0.43(3) | 0.60(3) | 0.70(2) | 0.93(5) | 1.8(2) | | | |
| $l_{blk} = 1024$ | 0.452(3) | 0.643(3) | 0.744(3) | 0.906(4) | 1.22(2) | | | |
| $l_{blk} = 2048$ | 0.456(2) | 0.644(2) | 0.747(3) | 0.902(4) | 1.22(2) | | | • |
| $l_{blk} = 4096$ | 0.458(3) | 0.646(2) | 0.749(3) | 0.900(4) | 1.21(2) | | | |
| $l_{blk} = 8192$ | 0.457(2) | 0.649(3) | 0.749(3) | 0.898(5) | 1.21(2) | | | |
| $\bar{r} = 4$ | 0.455(2) | 0.646(2) | 0.750(3) | 0.991(6) | | | | |
| $\bar{r} = 5$ | 0.456(2) | 0.644(2) | 0.747(3) | 0.902(4) | 1.22(2) | | | • |
| $\bar{r} = 6$ | 0.455(2) | 0.644(3) | 0.745(3) | 0.896(4) | 1.022(4) | 1.44(4) | | |
| $\bar{r} = 7$ | 0.455(2) | 0.643(3) | 0.744(4) | 0.895(4) | 1.019(4) | 1.141(5) | 1.50(5) | |
| ① | 0.459(1) | 0.650(3) | 0.748(4) | 0.899(4) | 1.200(9) | | | |
| ② | 0.455(1) | 0.645(2) | 0.744(2) | 0.899(3) | 1.22(2) | | | |
| ③ | 0.456(2) | 0.644(2) | 0.747(3) | 0.902(4) | 1.22(2) | | | • |
| $24^3 \cdot 32$ | 0.456(2) | 0.644(2) | 0.747(3) | 0.902(4) | 1.22(2) | | | • |
| $24^3 \cdot 60$ | 0.454(2) | 0.652(2) | 0.742(3) | 0.909(3) | 1.23(2) | | | |
| pl. waves | 0.454(2) | 0.647(2) | 0.751(4) | 0.905(4) | 1.028(4) | | | |
| Lüscher fct. | 0.456(2) | 0.644(2) | 0.747(3) | 0.902(4) | 1.22(2) | | | • |
| $W_\nu \to 4\sinh\frac{W_\nu}{4}$ | 0.457(2) | 0.647(3) | 0.751(3) | 0.909(4) | 1.24(2) | | | |

**Table 6**: Comparison of the two-particle energy data in the isospin-0 channel at ($\kappa = 0.315$, $J = 0.01$) for different methods of analysis: different reference times $t_0$, block lengths $l_{blk}$ and dimensions $\bar{r}$ of the correlation function matrix. Further variations concern the choice of the wave function and the strategy for calculating the block mean values (fig. 6). Additionally we compare lattices with different time extent: $24^3 \cdot 32$ and $24^3 \cdot 60$ lattices. The lines with • refer to our choice of final results (they are displayed several times).

| $\kappa = 0.310$ |
|---|
| $J = 0.005$ |

## two-particle energy spectrum

### isospin 0

| | $W_0$ | $W_1$ | $W_2$ | $W_3$ | $W_4$ | $W_5$ | $W_6$ |
|---|---|---|---|---|---|---|---|
| $16^3 \cdot 32$ | 0.352(2) | 0.548(2) | 1.032(8) | | | | |
| | 0.363 | 0.548 | 0.903 | | | | |
| $20^3 \cdot 32$ | 0.358(2) | 0.526(3) | 0.753(2) | 1.15(2) | | | |
| | 0.368 | 0.531 | 0.759 | 0.99 | | | |
| $24^3 \cdot 32$ | 0.360(3) | 0.520(3) | 0.661(3) | 0.849(4) | 1.28(5) | | |
| | 0.371 | 0.519 | 0.669 | 0.853 | 0.99 | | |
| $32^3 \cdot 40$ | 0.368(2) | 0.493(2) | 0.566(3) | 0.681(2) | 0.780(5) | 1.12(5) | |
| | 0.373 | 0.498 | 0.570 | 0.689 | 0.785 | 0.87 | |

### isospin 2

| | $W_0$ | $W_1$ | $W_2$ | $W_3$ | $W_4$ | $W_5$ |
|---|---|---|---|---|---|---|
| $16^3 \cdot 32$ | 0.368(1) | 0.871(1) | 1.165(2) | | | |
| | 0.376 | 0.877 | 1.182 | | | |
| $20^3 \cdot 32$ | 0.369(2) | 0.730(1) | 0.962(1) | 1.142(2) | | |
| | 0.375 | 0.736 | 0.971 | 1.154 | | |
| $24^3 \cdot 32$ | 0.363(2) | 0.641(1) | 0.829(1) | 0.973(1) | 1.091(1) | |
| | 0.375 | 0.646 | 0.834 | 0.984 | 1.114 | |
| $32^3 \cdot 40$ | 0.369(1) | 0.538(1) | 0.667(1) | 0.771(1) | 0.861(1) | 0.949(1) |
| | 0.374 | 0.544 | 0.672 | 0.778 | 0.871 | 0.958 |

**Table 7**: The two-particle energy spectrum for $\kappa = 0.310$ and $J = 0.005$ calculated as described in sect. 6.5. For comparison we also show the perturbative predictions (numbers without errors) based on the results of table 11. Only some higher energy levels above the four particle threshold $W > 4m_\pi = 0.748(3)$ (indicated by the zigzag line) show larger deviations. The errors given are purely statistical.

| |
|---|
| $\kappa = 0.315$<br>$J = 0.010$ |

## two-particle energy spectrum

### isospin 0

| | $W_0$ | $W_1$ | $W_2$ | $W_3$ | $W_4$ | $W_5$ | $W_6$ |
|---|---|---|---|---|---|---|---|
| $16^3 \cdot 32$ | 0.447(3) | 0.703(4) | 0.948(4) | 1.36(2) | | | |
| | 0.454 | 0.704 | 0.953 | 1.24 | | | |
| $20^3 \cdot 32$ | 0.454(3) | 0.673(3) | 0.814(4) | 1.024(4) | 1.41(4) | | |
| | 0.457 | 0.678 | 0.822 | 1.035 | 1.20 | | |
| $24^3 \cdot 32$ | 0.456(2) | 0.644(2) | 0.747(3) | 0.902(4) | 1.22(2) | | |
| | 0.458 | 0.651 | 0.748 | 0.904 | 1.030 | | |
| $32^3 \cdot 40$ | 0.450(2) | 0.588(2) | 0.667(2) | 0.753(2) | 0.834(2) | 0.912(3) | 1.21(2) |
| | 0.459 | 0.593 | 0.676 | 0.760 | 0.835 | 0.916 | 1.01 |

### isospin 2

| | $W_0$ | $W_1$ | $W_2$ | $W_3$ | $W_4$ | $W_5$ |
|---|---|---|---|---|---|---|
| $16^3 \cdot 32$ | 0.457(3) | 0.905(2) | 1.193(3) | 1.413(3) | | |
| | 0.461 | 0.916 | 1.211 | 1.441 | | |
| $20^3 \cdot 32$ | 0.458(2) | 0.776(2) | 0.994(1) | 1.168(3) | 1.306(3) | |
| | 0.460 | 0.782 | 1.006 | 1.184 | 1.340 | |
| $24^3 \cdot 32$ | 0.458(2) | 0.694(1) | 0.865(1) | 1.012(2) | 1.127(2) | 1.240(2) |
| | 0.460 | 0.699 | 0.875 | 1.019 | 1.145 | 1.263 |
| $32^3 \cdot 40$ | 0.453(1) | 0.602(1) | 0.720(1) | 0.817(1) | 0.901(1) | 0.987(1) |
| | 0.460 | 0.606 | 0.723 | 0.822 | 0.911 | 0.994 |

**Table 8**: The two-particle energy spectrum for $\kappa = 0.315$ and $J = 0.01$ calculated as described in sect. 6.5. For comparison we also show the perturbative predictions (numbers without errors) based on the results of table 11. Only some higher energy levels above the four particle threshold $W > 4m_\pi = 0.920(3)$ (indicated by the zigzag line) show larger deviations. The errors given are purely statistical.

| $\kappa = 0.315$ |
|---|
| $J = 0.010$ |

**two-particle energy spectrum**

**isospin 1**

| | $W_0$ | $W_1$ | $W_2$ | $W_3$ | $W_4$ |
|---|---|---|---|---|---|
| $16^3 \cdot 32$ | 0.897(2) | 1.181(3) | 1.401(2) | | |
| | 0.909 | 1.201 | 1.435 | | |
| $20^3 \cdot 32$ | 0.769(1) | 0.988(2) | 1.166(1) | 1.303(2) | |
| | 0.778 | 1.000 | 1.181 | 1.340 | |
| $24^3 \cdot 32$ | 0.690(1) | 0.863(1) | 1.003(2) | 1.121(2) | 1.233(2) |
| | 0.697 | 0.871 | 1.016 | 1.144 | 1.257 |
| $32^3 \cdot 40$ | 0.599(1) | 0.714(1) | 0.814(1) | 0.901(1) | 0.980(1) |
| | 0.605 | 0.721 | 0.821 | 0.910 | 0.991 |

**Table 9**: The two-particle energy spectrum for $\kappa = 0.315$, $J = 0.01$ and isospin 1 calculated as described in sect. 6.5. For comparison we also show the perturbative predictions (numbers without errors) based on the results of table 11. Only some higher energy levels above the four particle threshold $W > 4m_\pi = 0.920(3)$ (indicated by the zigzag line) show larger deviations. The errors given are purely statistical.

| $\kappa = 0.320$<br>$J = 0.020$ |
|---|

## two-particle energy spectrum

### isospin 0

| | $W_0$ | $W_1$ | $W_2$ | $W_3$ | $W_4$ | $W_5$ | $W_6$ |
|---|---|---|---|---|---|---|---|
| $12^3 \cdot 32$ | 0.575(2) | 0.888(4) | 1.240(3) | 1.80(2) | | | |
| | 0.579 | 0.894 | 1.261 | 1.652 | | | |
| $16^3 \cdot 32$ | 0.582(4) | 0.844(5) | 1.031(5) | 1.29(1) | 1.79(6) | | |
| | 0.585 | 0.852 | 1.038 | 1.30 | 1.50 | | |
| $20^3 \cdot 32$ | 0.582(3) | 0.802(4) | 0.920(4) | 1.096(5) | 1.248(6) | 1.60(6) | |
| | 0.587 | 0.810 | 0.928 | 1.109 | 1.256 | 1.40 | |
| $24^3 \cdot 40$ | 0.582(3) | 0.756(3) | 0.861(3) | 0.977(4) | 1.101(8) | 1.192(5) | 1.51(6) |
| | 0.588 | 0.767 | 0.869 | 0.992 | 1.100 | 1.209 | 1.33 |

### isospin 2

| | $W_0$ | $W_1$ | $W_2$ | $W_3$ | $W_4$ | $W_5$ |
|---|---|---|---|---|---|---|
| $12^3 \cdot 32$ | 0.587(2) | 1.189(1) | 1.558(2) | 1.842(3) | | |
| | 0.591 | 1.211 | 1.607 | 1.914 | | |
| $16^3 \cdot 32$ | 0.589(2) | 0.978(1) | 1.246(2) | 1.449(2) | 1.608(3) | |
| | 0.590 | 0.986 | 1.265 | 1.487 | 1.680 | |
| $20^3 \cdot 32$ | 0.587(3) | 0.857(2) | 1.059(2) | 1.221(2) | 1.353(2) | 1.487(3) |
| | 0.590 | 0.864 | 1.071 | 1.240 | 1.390 | 1.529 |
| $24^3 \cdot 40$ | 0.586(2) | 0.786(2) | 0.940(1) | 1.074(2) | 1.185(2) | 1.292(2) |
| | 0.590 | 0.790 | 0.949 | 1.083 | 1.203 | 1.315 |

**Table 10**: The two-particle energy spectrum for $\kappa = 0.320$ and $J = 0.020$ calculated as described in sect. 6.5. For comparison we also show the perturbative predictions (numbers without errors) based on the results of table 11. Only some higher energy levels above the four particle threshold $W > 4m_\pi = 1.179(3)$ (indicated by the zigzag line) show larger deviations. The errors given are purely statistical.

# 8 The properties of the wave functions

For both kinds of wave functions (6.3) and (6.4) we obtain the same energy spectrum within a standard deviation. This fact shows how reliable our diagonalization procedure is. However, the question remains, if Lüscher's wave functions (6.4) have an improved projection on energy eigenstates compared with the plane waves (6.3).

In order to get an idea of the projection properties we consider the normalized spectral amplitudes $c_j^\nu$ (cf. eq. (6.15)). In figs. 11 and 12 we plot $|c_j^\nu|^2$ for the isospin-0 operators $O_j, j = \sigma, 0, 1, \ldots, \bar{r}-2$, with plane waves and Lüscher's wave functions, respectively, at ($\kappa = 0.315$, $J = 0.01$, $L = 24$). Each box shows $|c_j^\nu|^2$ for a fixed value of $j$. Within statistical errors, the results for $|c_\sigma^\nu|^2$ are independent of the type of wave function used in the operators $O_0, O_1, \ldots$ Of course, this has to be so, since $O_\sigma$ is the same operator in both cases. This agreement provides another consistency check of our calculations. In the case of the plane waves we see clearly why it was vital to include the $\sigma$-operator: $O_0, O_2, O_3$ and $O_4$ generate essentially only the energy eigenstates $|\nu\rangle$ with $\nu = 0, 3, 4, 5$. So $O_1$ needs a partner, if $|1\rangle$ is to be separated from $|2\rangle$.

The operators $O_j$ constructed with Lüscher's wave functions do what they are supposed to do: $O_j$ generates predominantly the energy eigenstate $|j\rangle$, except that $O_1$ leads to a strong $\nu = 2$ component also. This could be due to an inaccurate estimate of $q_1^2$, which had to be fixed beforehand (cf. sect. 6.1). In summary we have to conclude that there is no essential difference between Lüscher's wave functions and plane waves.

# 9 Numerical results for the scattering phases

For each lattice extent $L$ and each two-pion energy level $W_\nu < 4m_\pi$ we get one value of the scattering phase shift $\delta_0^0$, $\delta_1^1$, and $\delta_0^2$, respectively, in the elastic region $0 < k_\nu/m_\pi < \sqrt{3}$ using the key relation (2.4)

$$\delta_l^I(k_\nu) = -\phi\left(\frac{k_\nu L}{2\pi}\right) \bmod \pi \,. \tag{9.1}$$

The momentum $k_\nu$ corresponding to $W_\nu$ has to be calculated with the help of the energy momentum relation (6.17) as discussed in sect. 6.3.

Fig. 13 illustrates the calculation of the scattering phase shifts for the simulation point ($\kappa = 0.315$, $J = 0.01$). For some fixed values of $m_\pi L$ we plot the function $-\phi\left(\frac{kL}{2\pi}\right) \bmod \pi$, already shown in fig. 1, versus $k/m_\pi$. The value of this function at a momentum $k_\nu(W_\nu) < \sqrt{3}\, m_\pi$ gives the scattering phase shift at that momentum. The data points in fig. 13 show the momenta extracted from the two-pion energy levels and the corresponding phase shifts whose errors are determined by the slope of $\phi$. Each passing of the function $-\phi$ through a multiple of $\pi$ turns out to correspond to one energy level $W_\nu$.

Comparing the different values of $L$ we observe the following: As $L$ increases, more and more energy levels lie below the inelastic threshold. At the same time the function $-\phi\left(\frac{kL}{2\pi}\right)$ becomes steeper when considered as function of $k$. This fact has important consequences for the error propagation: On larger lattices higher accuracy of the energy values $W_\nu$ is required, though it is increasingly harder to achieve.

There is another consequence of the steepness of the function $-\phi$. The difference between the energy values determined with or without lattice effects taken into account is in general not much larger than the statistical error. Especially at the simulation point ($\kappa = 0.320$, $J = 0.02$), where the largest energy values appear, this difference leads to a systematic shift of the resulting scattering phases due to the large slope of the function $-\phi$. Fig. 14 illustrates this effect: Using the lattice dispersion relation (6.17) instead of the continuum dispersion relation the scattering phases are shifted downwards along the dotted lines representing $-\phi\left(\frac{k_\nu L}{2\pi}\right) \bmod \pi$.

How do our results compare with perturbation theory? In fig. 15 we plot all values for $\delta_0^0$ that we obtained at ($\kappa = 0.315$, $J = 0.01$) together with the perturbative prediction based on the estimates $\tilde{m}_\sigma$ and $\tilde{g}_R$ (dashed curve). The comparison shows that the value of the resonance mass $m_\sigma$ lies below the estimate $\tilde{m}_\sigma$ from the fit to the $\sigma$-propagator in momentum space.

Figs. 17 and 16 summarize our results for all isospin values and simulation points. For isospin 1 and 2 the agreement of the numerical results and the perturbative prediction is quite good (even if we use the estimates $\tilde{m}_\sigma$ and $\tilde{g}_R$ of table 4). In these cases the use of the lattice dispersion relation (6.17) is essential. Note that the data fall onto a single curve, although they originate from different lattice sizes. This justifies *a posteriori* the assumption that the influence of higher angular momenta and finite volume polarization effects can be neglected.

Finally we discuss different methods for the determination of the resonance mass $m_\sigma$ and the decay width $\Gamma_\sigma$ from the measured scattering phases:

1. **Breit-Wigner-fit (model independent):** Fitting the values of isospin-0 scattering phases around the resonance ($\pi/4 \lesssim \delta_0^0 \lesssim 3\pi/4$) by means of the (relativistic) Breit-Wigner formula
$$\tan\Big(\delta_0^0 - \frac{\pi}{2}\Big) = \frac{W^2 - m_\sigma^2}{m_\sigma \Gamma_\sigma} \tag{9.2}$$
it is possible to calculate the resonance mass $m_\sigma$ and its width $\Gamma_\sigma$. One major advantage of this fit is the determination of the resonance parameters $m_\sigma$ and $\Gamma_\sigma$ without additional assumptions concerning the model under investigation. The fit works reasonably well although there are only few data points inside the resonance region ($\pi/4 \lesssim \delta_0^0 \lesssim 3\pi/4$).
2. **Perturbative fits (model dependent):** Outside the resonance region the Breit-Wigner-formula (9.2) does not apply. Using the perturbative formula (3.6)
$$\delta_0^0 = \delta_{0,r}^0 + \delta_{0,s}^0 \tag{9.3}$$
$$\text{with} \qquad \tan\delta_{0,s}^0 = g_R \frac{N-1}{48\pi}\,\frac{m_\sigma^2 - m_\pi^2}{m_\sigma^2 - W^2}\,\frac{k}{W}$$
$$\text{and} \qquad \delta_{0,r}^0 = \frac{g_R}{96\pi}\,\frac{m_\sigma^2 - m_\pi^2}{kW}\ln\Big(\frac{4k^2 + m_\sigma^2}{m_\sigma^2}\Big) - g_R\frac{N+1}{48\pi}\,\frac{k}{W}$$
it is possible to fit all data points in the elastic region by a two-parameter fit with respect to the resonance mass $m_\sigma$ and the coupling $g_R$.
The width $\Gamma_\sigma$ of the $\sigma$-resonance may then be calculated from the perturbative formula [21]:
$$\Gamma_\sigma = g_R \frac{N-1}{96\pi}\,\frac{m_\sigma^2 - m_\pi^2}{m_\sigma^2}\,\sqrt{m_\sigma^2 - 4m_\pi^2}\,. \tag{9.4}$$
If in addition to the pion mass $m_\pi$ also the values of the wave function renormalization constant $Z$ and the field expectation value $\Sigma$ are taken to be known from table 4, the same ansatz (9.3) can be used in order to perform a 1-parameter fit with respect to $m_\sigma$ provided that we replace $g_R$ by its definition
$$g_R = 3Z\frac{(m_\sigma^2 - m_\pi^2)}{\Sigma^2}\,. \tag{9.5}$$
The results of the latter fit serve as basis for the perturbative calculations of the energy spectrum (see tables 7 to 10) and the computation of the curves shown in our figures.

The results for the resonance mass $m_\sigma$, the decay width $\Gamma_\sigma$ and the coupling $g_R$ obtained by means of these fits are summarized in table 11. If one of the variables $g_R$ or $\Gamma_\sigma$ is not a fit parameter it is calculated by means of the formulæ (9.5) and (9.4), respectively. The values do not depend very much on the kind of fit ansatz used. The results for the decay width agree reasonably well with the perturbative predictions, while the resonance masses $m_\sigma$ lie systematically (about 5 %) below the estimates $\tilde{m}_\sigma$ from the fit to the propagator in momentum space. However, this discrepancy should not be too surprising, because the form of the propagator used in the determination of $\tilde{m}_\sigma$ corresponds to a stable particle and the width of the $\sigma$-resonance for our choice of parameters is rather large: $\Gamma_\sigma \approx 0.11$–$0.18\, m_\sigma$.

| | SIMULATION POINTS | | |
|---|---|---|---|
| | **1** | **2** | **3** |
| $\kappa$ | 0.310 | 0.315 | 0.320 |
| $J$ | 0.005 | 0.010 | 0.020 |
| **Estimates** | | | |
| $\tilde{m}_\sigma$ | 0.538(1) | 0.720(1) | 0.906(2) |
| $\tilde{g}_R(\tilde{m}_\sigma, m_\pi, \Sigma, Z)$ | 22.0(1) | 24.4(1) | 27.0(1) |
| $\tilde{\Gamma}_\sigma(\tilde{m}_\sigma, m_\pi, \tilde{g}_R)$ | 0.074(1) | 0.121(1) | 0.165(1) |
| **Breit-Wigner-fit** | | | |
| $m_\sigma$ | 0.520(1) | 0.706(2) | 0.885(2) |
| $g_R(m_\sigma, m_\pi, \Sigma, Z)$ | 20.3(1) | 23.4(1) | 25.8(2) |
| $\Gamma_\sigma$ | 0.079(4) | 0.130(9) | 0.160(17) |
| **Perturbative fit to $m_\sigma$ & $g_R$** | | | |
| $m_\sigma$ | 0.513(1) | 0.688(2) | 0.863(3) |
| $g_R$ | 21.1(8) | 25.7(1.0) | 28.1(1.4) |
| $\Gamma_\sigma(m_\sigma, m_\pi, g_R)$ | 0.064(3) | 0.116(5) | 0.156(8) |
| **Perturbative fit to $m_\sigma$** | | | |
| $m_\sigma$ | 0.515(1) | 0.691(1) | 0.868(2) |
| $g_R(m_\sigma, m_\pi, \Sigma, Z)$ | 19.8(1) | 22.3(1) | 24.6(1) |
| $\Gamma_\sigma(m_\sigma, m_\pi, g_R)$ | 0.060(1) | 0.102(1) | 0.138(1) |

**Table 11**: Results for the resonance mass $m_\sigma$, the coupling $g_R$ and the decay width $\Gamma_\sigma$ of the $\sigma$-particle obtained by fits with the relativistic Breit-Wigner-formula (9.2) and by one- or two-parameter fits with the perturbative ansatz (3.6), respectively. If the coupling $g_R$ or the decay width $\Gamma_\sigma$ are not fit parameters, they are calculated by means of eqs. (9.5) and (9.4), respectively. These results are compared with the estimates of table 4. The errors given are purely statistical.

# 10 Conclusion

The main result of our paper is that Lüscher's method for studying particle scattering in massive quantum field theories on finite lattices can be applied successfully even in four dimensions. It is numerically very stable, i.e. the results vary only within statistical errors under a change of the parameters of the analysis (see sect. 6.3) and it passes all our selfconsistency checks (see e.g. sect. 8). Furthermore it was confirmed *a posteriori* that correction terms to the key relation (2.4) resulting from polarisation effects and contributions of higher angular momenta can be neglected in our simulations (see sect. 7). Scaling violations seem to be compensated (at least partially) by using the lattice dispersion relation instead of the continuum dispersion relation, as already observed for single-particle states.

In the isospin-0 channel the expected $\sigma$-resonance is found. Its mass and width can be extracted with reasonable accuracy. Measuring the mass by a fit to the propagator in momentum space gives – not surprisingly – a slightly different value. The discrepancy is, however, small enough to leave the previous calculations of an upper bound to the Higgs boson mass essentially unchanged (unless the difference increases considerably in the limit $J \rightarrow 0$).

The other channels show no resonance structure. In accordance with [28] we see no $\rho$-resonance. Hence the formation of the $\rho$-meson cannot be understood in the framework of the $\Phi^4$ model. However, this conclusion has to be considered with some caution. If the $\sigma$-operator is omitted in the study of the isospin-0 channel, all signs of a resonance disappear. Could it be that we have left out the "essential" operator in the isospin-1 channel?

Since one might argue that an operator which is more strongly localized than the plane waves used above should have an enhanced projection on the resonance, we have added in the analysis of the isospin-1 channel the operator

$$\frac{1}{L^3} \sum_{\vec{x} \in \mathbb{Z}_L^3} \epsilon^{abc} \Phi^a(\vec{x}, t) \Phi^b(\vec{x} + \vec{e}_j, t) , \tag{10.1}$$

which corresponds to the conserved $O(3)$-current in the continuum limit. But its inclusion does not alter our results.

As already observed in other investigations, due to the triviality of the $\Phi^4$-theory in four dimensions renormalized perturbation theory is reliable. We find complete agreement with our nonperturbative results, provided the renormalized parameters, especially $m_\sigma$, are chosen properly. This finding corroborates our above statement that Lüscher's method works well. Furthermore, in view of the agreement with perturbation theory, it seems unlikely to us that we have overlooked a resonance in the isospin-1 channel, which would be a nonperturbative phenomenon.

The successful application of Lüscher's method to a four-dimensional quantum field theory as well as the above mentioned investigations of lower dimensional systems give rise to some optimism concerning the study of more demanding models, in particular QCD.

# Acknowledgement

Helpful discussions with M. Lüscher and C. Frick are gratefully acknowledged. Special thanks are due to T. Neuhaus for allowing us to use his programs as basis of our programming. Furthermore we wish to thank the *Rechenzentrum* at the RWTH Aachen and the HLRZ Jülich for providing the necessary computer time.